\documentclass[iop]{emulateapj}
\slugcomment{{\sc Accepted by ApJ:} May 5 2012}
\pdfoutput=1
\usepackage{graphicx}
\usepackage{natbib}
\usepackage{amsmath}
\usepackage{color}

\newcommand{\refeq}[1]{(\ref{#1})}
\newcommand{\stx}[1]{{*,\mathrm{#1}}}
\newlength{\floatwidth}
\begin{document}
\title{Shapes and Probabilities of Galaxy Clusters II: Comparisons with observations}
\author{Abel Yang}
\affil{Department of Astronomy, University of Virginia, Charlottesville, VA 22904}
\and
\author{William C. Saslaw}
\affil{Institute of Astronomy, Madingley Road, Cambridge CB3 0HA, UK; and Department of Astronomy, University of Virginia, Charlottesville, VA 22904}
\begin{abstract}
We identify low redshift clusters and groups in the Sloan Digital Sky Survey (SDSS) and estimate their kinetic and correlation potential energies. We compare the distribution of these energies to the predictions by \citet{2012ApJ...745...87Y} and in the process estimate a measure of an average 3-dimensional velocity and spatial anisotropy of a sample of clusters. We find that the inferred velocity anisotropy is correlated with the inferred spatial anisotropy. We also find that the general shape of the energy distribution agrees with theory over a wide range of scales from small groups to superclusters once the uncertainties and fluctuations in the estimated energies are included.
\end{abstract}


\defcitealias{2012ApJ...745...87Y}{paper~1}

\section{Introduction}

Clusters and groups of galaxies are structures in the universe which have been defined using various criteria; groups are smaller clusters. These clusters may contain over a thousand member galaxies and occupy a few cubic megaparsecs. Although clusters can be identified by their density and size~(e.g. \citealt{1958ApJS....3..211A, 1957PASP...69..409H}) their shapes and structures differ. For example, \citet{1957PASP...69..409H} identify three classes of clusters: \textit{compact clusters} which have a single nearly spherical dense concentration of galaxies, \textit{medium compact clusters} which are less dense and may have multiple concentrations of galaxies and \textit{loose clusters} which do not have any outstanding concentrations of galaxies.

Even these simple classes of clusters suggest greater complexity than just spherical concentrations in a region of space. For example, \citet{1987AJ.....94..251B} found significant substructure in the core of the Virgo cluster as well as pronounced double structure, and \citet{1957PASP...69..409H} classify the Virgo cluster as a medium compact cluster. The structure of the Virgo cluster, as the nearest large cluster, suggests that these irregular shapes are common.

Such irregular shapes may result from mergers. Smaller groups fall into the central region of a cluster and form subgroups whose member galaxies are still tightly bound to each other. Irregular shapes resulting from subgroups then disappear as a cluster virializes. However, many clusters have dynamical relaxation timescales on the order of a Hubble time, and their incomplete virialization suggests that irregular clusters with multiple concentrations of galaxies should be common in the universe.

The basic dynamical description of a cluster is its 6-dimensional phase space configuration such as a sphere with a density and velocity profile. More detailed descriptions of clustering include correlation functions, percolation trees and counts-in-cells statistics. In particular, the counts-in-cells description is especially suitable for this problem because it straightforwardly analyzes regions of space~(cells) with a specified size and shape. In addition, the physics of this description can be derived from gravitational thermodynamics~\citep{1984ApJ...276...13S} or statistical mechanics~\citep{2002ApJ...571..576A} where the galaxies are in quasi-equilibrium and interact in a grand canonical ensemble of cells.

While self-gravitating systems, and thus cells, are not in strict equilibrium, they are in quasi-equilibrium when the average energies and thermodynamic quantities of an ensemble of cells change slowly compared to the dynamical timescale of a single cell. This means that the intermediate time averages of the ensemble are stable, while the local snapshot energies of a cell fluctuate about their quasi-equilibrium time averages. The quasi-equilibrium approximation thus allows us to study self gravitating systems using thermodynamics originally intended for systems in equilibrium. 

The resulting counts-in-cells distribution is thus known as the gravitational quasi-equilibrium distribution~(GQED), The simplest form of the counts-in-cells GQED is~\citep{2000dggc.book.....S}
\begin{equation}\label{eq-fvn}
f_V(N) = \frac{\overline{N}(1-b)}{N!}\left[\overline{N}(1-b)+N b\right]^{N-1} e^{-\overline{N}(1-b)-Nb}
\end{equation}
which describes the probability that a cell of volume $V$ has $N$ galaxies. This depends on the average number of galaxies in a cell $\overline{N}$ and a clustering parameter $b$ which is related to the mean and variance of $f_V(N)$ through the dimensionless equation~\citep{2002ApJ...571..576A}
\begin{equation}\label{eq-bdefvar}
b = 1-\sqrt{\overline{N}/\langle(\Delta N)^2\rangle}.
\end{equation}
The GQED therefore describes the clustering of galaxies with no free parameters.

The implications of this physical description for the shapes of clusters was studied in \citet[paper 1]{2012ApJ...745...87Y}, which provided a method for determining the probability that a cell with $N$ galaxies has a particular kinetic energy and correlation potential energy. This method uses the statistical mechanics of the GQED to describe the probability that a cell has a given kinetic energy and correlation potential energy, and is based on earlier work by \citet{2004ApJ...608..636L}.

The manner in which the internal structures and shapes of clusters of galaxies are connected to the large scale structure of the universe through the GQED therefore provides an opportunity for an observational test of the theory in \citetalias{2012ApJ...745...87Y} and earlier work, as well as a means to study the internal structure of galaxy clusters. To do so, we analyze the New York University Value-added Galaxy Catalog~(NYU-VAGC) derived from the Sloan Digital Sky Survey~(SDSS).

While a simulation may be seem to be easier to analyze with few uncertainties, simulations are essentially approximations of the universe that involve other assumptions and uncertainties that simplify the problem. These include the choice of simulation volume, the mass resolution and the initial conditions. Their associated effects may introduce further complications in a poorly designed simulation. Therefore, this paper examines comparisons with observations which can determine an appropriate set of constraints and parameters for future simulations.

This paper is structured as follows: In section \ref{sec-theory} we describe the theoretical background with reference to earlier work by \citet{2012ApJ...745...87Y} which we refer to as \citetalias{2012ApJ...745...87Y}. In particular, we analyze the relation between the kinetic energy, gravitational correlation potential energy and total energy within the framework of the GQED. In section \ref{sec-data} we describe the NYU-VAGC samples from the SDSS and our selection cuts. In section \ref{sec-analysis} we describe the procedures we develop to obtain the counts-in-cells parameters $\overline{N}$ and $b$, and the algorithm we use to identify cells that contain clusters from the catalog. In section \ref{sec-comp} we compare our observations with the theory. Finally we discuss our results in section \ref{sec-conc}. In this paper we use $\Omega_m = 0.3$, $\Omega_k = 0.0$, $\Omega_\Lambda = 0.7$ and $H_0 = 100 h$ km s$^{-1}$ Mpc$^{-1}$ following \citet{2005AJ....129.2562B}.

\section{Theoretical Background}
\label{sec-theory}
The theoretical background of this paper is based on earlier work by \citet{2012ApJ...745...87Y} which we refer to as \citetalias{2012ApJ...745...87Y}. To begin, we describe the kinetic and correlation potential energies of a cell in terms of scaled dimensionless variables and factor out the dimensional quantities such as the cell radius and average mass of a galaxy. Thus we write distances in units of cell radii, time in units of dynamical times~(c.f. equation \ref{eq-tcrossest} below) and masses in terms of the average mass of a galaxy.

From this, we can write the observed scaled dimensionless energies of a cell with $N$ galaxies as~(Equations (61) and (69) of \citetalias{2012ApJ...745...87Y})
\begin{equation}\label{eq-Wsobs}
W_* = -\frac{4}{9}\frac{(N-1)}{N\zeta(\epsilon/R)}R\left\langle\frac{\kappa(r/\epsilon)}{r_{12}}\right\rangle
\end{equation}
for the scaled correlation potential energy and
\begin{equation}\label{eq-Tsobs}
T_* = \frac{4}{9}\frac{\upsilon^2}{\zeta(\epsilon/R)}
\end{equation}
for the scaled kinetic energy. Here $R$ is the radius of a cell and $\upsilon$ is the peculiar velocity of a galaxy in units of cell radii per dynamical time. These scaled energies are written in terms of the average potential energy of a galaxy in a cell so that $T_*$ represents a ratio of the kinetic energy of a cell to the average potential energy of a galaxy and is related to the correlation virial ratio, while $W_*$ is a measure of the compactness of the galaxy distribution within a cell.

In equations \refeq{eq-Wsobs} and \refeq{eq-Tsobs}, $\kappa(r)$ is a dimensionless modification factor to a point mass potential that describes its departure from a Newtonian point-source potential through
\begin{equation}\label{eq-kappadef}
\phi(r) = -\frac{Gm^2}{r}\kappa(r/\epsilon)
\end{equation}
where $\epsilon$ is a parameter that describes the strength of the modification. This modification may be caused by an extended halo~\citep{2002ApJ...571..576A} or a merging pair~\citep{2010arXiv1011.0176Y} among other possibilities. The $\zeta(\epsilon/R)$ factor is related to $\kappa(r/\epsilon)$ through~(equation (6) of \citetalias{2012ApJ...745...87Y})
\begin{equation}\label{eq-zetadef}
\zeta\left(\frac{\epsilon}{R}\right) = \frac{2}{R^2}\int_0^{R} r\kappa(r/ \epsilon)dr.
\end{equation}
and is of order unity when $\epsilon$ is small compared to the cell radius~(c.f. Figure 1 of \citealt{2002ApJ...571..576A}). In such cases, galaxies may be reasonably approximated as point masses.

Because these scaled energies come from observations, $T_*$ and $W_*$ describe a snapshot of the cell which fluctuates about its quasi-equilibrium values $\overline{T}_*$ and $\overline{W}_*$. These quasi-equilibrium energies are essentially time-averaged values of $T_*$ and $W_*$ taken over the fluctuation timescale of a cell. This is based on the property of the cosmological many-body problem that the macroscopic evolution of a region in quasi-equilibrium is slow~(approximately equal to or greater than a Hubble time, e.g. \citealt{2000dggc.book.....S}) compared to its crossing time so that equilibrium prevails approximately.

Regions that are not in quasi-equilibrium generally have a lower entropy than regions that are, and their configurations will usually change toward the higher-entropy quasi-equilibrium state~\citep[Section 15.2]{2000dggc.book.....S}. The timescale for this relaxation is approximately the dynamical crossing time of the configuration, and is shorter than the quasi-equilibrium evolution timescale of the entire system. This is because the quasi-equilibrium evolution timescale of the entire system is at least as long as a Hubble time, while clusters, being overdense regions, have a shorter crossing time. This suggests that quasi-equilibrium is a good approximation for statistically homogenous cosmological self-gravitating systems of galaxies at any epoch~\citep[Section 25.6]{2000dggc.book.....S}. The observed galaxy distribution strongly supports this assumption~(e.g. \citealt{2005ApJ...626..795S,2009ApJ...695.1121R,2011ApJ...729..123Y}).

Therefore, we emphasize the distinction between the snapshot and quasi-equilibrium~(time averaged) quantities because clusters of galaxies are not in strict equilibrium and their energies will fluctuate about their quasi-equilibrium value. This intrinsic fluctuation will mean that the observed snapshot energies of a specific cluster of galaxies may not agree with its quasi-equilibrium value, but the time-averaged energy of a cluster and the energies of an ensemble of clusters will be distributed about quasi-equilibrium.

To complete our description of the clustering of galaxies in a cell, we introduce two related quantities: The scaled correlation energy $E_* = W_*+T_*$ and the correlation virial ratio $\psi = -W_*/2T_*$. The quasi-equilibrium counterparts to these quantities are similarly defined as $\overline{E}_* = \overline{W}_* + \overline{T}_*$ and $\overline{\psi} = -\overline{W}_*/2\overline{T}_*$. In quasi-equilibrium, these quantities are further related to each other through~(\citealt{2004ApJ...608..636L} and \citetalias{2012ApJ...745...87Y})
\begin{equation}\label{eq-EsQE}
\overline{E}_* = \overline{T}_*\frac{\overline{T}_*^3-1}{\overline{T}_*^3+1},
\end{equation}
\begin{equation}\label{eq-WsQE}
\overline{W}_* = \overline{E}_* - \overline{T}_* = -\frac{2\overline{T}_*}{\overline{T}_*^3+1}
\end{equation}
and
\begin{equation}\label{eq-psiQE}
\overline{\psi} = -\frac{\overline{W}_*}{2\overline{T}_*} = \frac{1}{\overline{T}_*^3+1}.
\end{equation}

These energies are also related to the larger ensemble through the virial ratio. In particular, the clustering parameter can also be written as~(e.g. \citealt{1996ApJ...460...16S})
\begin{equation}\label{eq-bdefvir}
b = -\frac{W}{2K}
\end{equation}
where $W$ is the ensemble average correlation potential energy and $K$ is the ensemble average kinetic energy. While equation \refeq{eq-bdefvir} suggests that $b$ and $\overline\psi$ are similar, they describe very different systems. Here, $\overline\psi$ describes the quasi-equilibrium correlation virial ratio for a single cell, while $b$ describes the ensemble average over all cells in the ensemble. In particular, there may be cells that are virialized, with a value of $\overline\psi$ close to $1$, while other cells may have a lower value of $\overline\psi$. The average value of $\overline\psi$ taken over all cells in the ensemble is $b$ and the distribution of $\overline\psi$, which we discuss in the rest of this section, is closely related to the distribution of cell energies $P(E,N)$.

\subsection{Probabilities}
The probability that a cell with $N$ galaxies in quasi-equilibrium in a grand canonical ensemble has total energy $E$ is given by the usual result from statistical mechanics~(e.g. \citealt{2004ApJ...608..636L})
\begin{equation}\label{eq-pE1}
P(E, N) dE = g(E) \frac{e^{-E/T_0}e^{N\mu/T_0}}{Z_G} dE
\end{equation}
where $g(E)$ is the density of states having energy $E$, and $T_0$, $\mu$ and $Z_G$ are the temperature, chemical potential and partition function of the grand canonical ensemble. Here we use units of temperature where the Boltzmann constant is unity so temperature has energy units. This relates the scaled energies to the GQED.

To write equation \refeq{eq-pE1} in terms of the scaled energies, we use results described in \citetalias{2012ApJ...745...87Y}. The first is to project the grand canonical ensemble into a canonical ensemble using
\begin{equation}\label{eq-pE1N}
\begin{split}
P_N(E) dE &= f_V(N) P(E|N) dE \\
 &= f_V(N) g(E) \frac{e^{-E/T_0}e^{N\mu/T_0}}{Z_G} dE
\end{split}
\end{equation}
where $P(E|N) dE$ is the conditional probability that a cell has energy $E$ given that it has $N$ galaxies, and $f_V(N)$ is the counts-in-cells distribution of equation \refeq{eq-fvn}. Then, we use the entropy of a canonical ensemble of galaxies in quasi-equilibrium~\citep{2002ApJ...571..576A} to obtain the density of states in terms of $\overline{T}_*$
\begin{equation}\label{eq-gE2}
\begin{split}
&g(\overline{E}_*[\overline{T}_*]) = \frac{d\Omega}{d\overline{T}_*}\frac{d\overline{T}_*}{d\overline{E}_*} \\
&= \frac{3N}{2\overline{T}_*}\left[\frac{V}{N}\left(1+\overline{T}_*^{-3}\right)\right]^N \left(\frac{2 \pi m T}{\Lambda^2}\right)^{\frac{3N}{2}} e^{\frac{5N}{2}-\frac{3N}{1+\overline{T}_*^3}}
\end{split}
\end{equation}
where $\Omega$ is the number of energy states for a small range of energy $\Delta E$ so that $\Omega = g(E)\Delta E$.

Using equation \refeq{eq-gE2}, and the fugacity $e^{N \mu /T_0}$ and the grand canonical partition function $Z_G$ from \citet{2002ApJ...571..576A} to write equation \refeq{eq-pE1} explicitly, we get~(c.f. equation (33) of \citetalias{2012ApJ...745...87Y})
\begin{equation}\label{eq-pE}
\begin{split}
&P(\overline{E}_*[\overline{T}_*]|N) d\overline{E}_* = \frac{3N}{2\overline{T}_*}\left(\frac{\overline{N}}{N}\sqrt{\frac{\overline{T}_*^3 b}{1-b}}\right)^N\\
 &\times (1+\overline{T}_*^{-3})^N (1-b)^N \exp\Biggl[(N - \overline{N})(1-b)\\
 & + \frac{3N}{2}\frac{\overline{T}_*^3-1}{\overline{T}_*^3+1}  \left(1-\left[\frac{\overline{T}_*^3 b}{1-b}\right]^{\frac{1}{3}}\right)\Biggr] d\overline{E}_*
\end{split}
\end{equation}
which describes the probability that a cell with $N$ galaxies has a quasi-equilibrium scaled correlation energy of $\overline{E}_*$. This depends on the scaled quasi-equilibrium kinetic energy $\overline{T}_*$, the mean number of galaxies in a cell $\overline{N}$ and the clustering parameter $b$. This dependence on $\overline{N}$ and $b$ relates the internal structure of a cell to the large scale structure of the universe.

\subsection{Normalization}

To normalize the probability in equation \refeq{eq-pE}, we consider the range of $\overline{T}_*$ that represents quasi-equilibrium. In the limit of weak gravitational interactions, $\overline{T}_* \to \infty$ and the system approximates an ideal gas which is a limit of the GQED. For the case of small $\overline{T}_*$, we use the condition that in virial equilibrium, the crossing time of a cell is approximately its dynamical time. This gives us a minimum value of $\overline{T}_\stx{min} \approx 0.1$. We therefore use $\overline{T}_\stx{min} = 0.1$ following \citetalias{2012ApJ...745...87Y} and normalize the probabilities for the range $0.1 \leq \overline{T}_* \leq \infty$.

\begin{figure*}[tb]
\begin{center}
\includegraphics[width=\floatwidth]{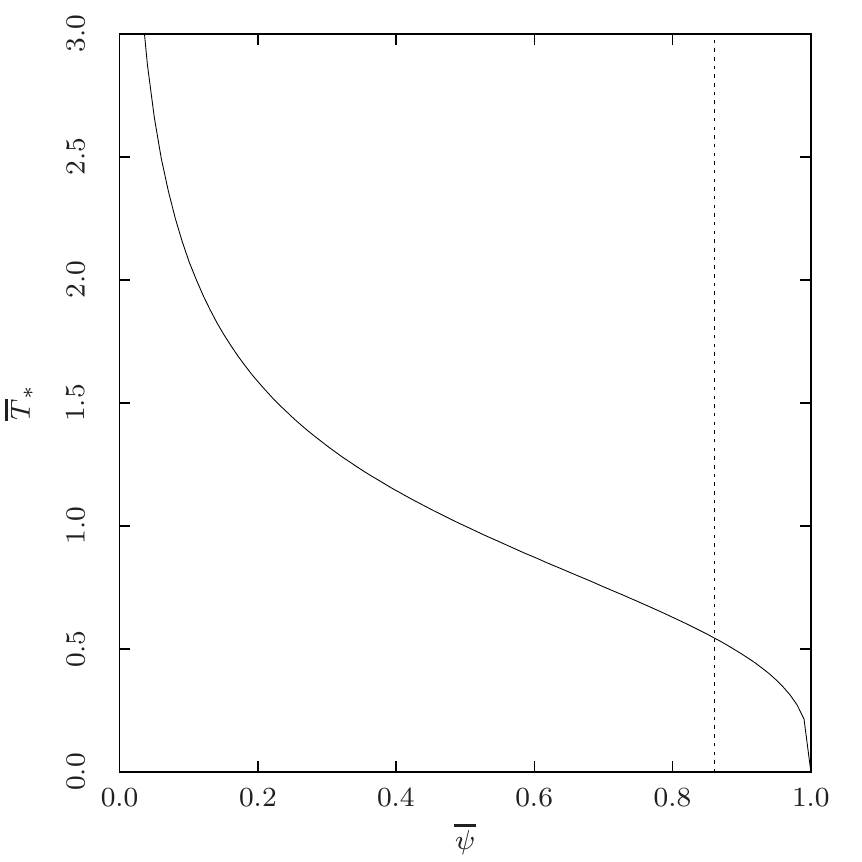}
\includegraphics[width=\floatwidth]{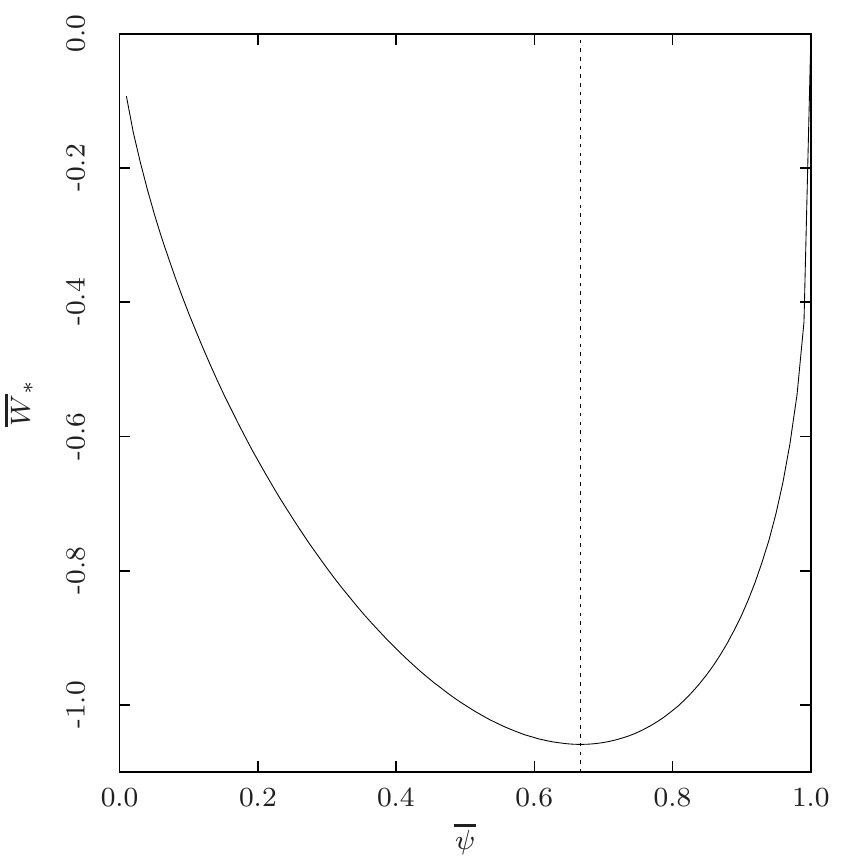}
\caption{
Left plot: $\overline{T}_*$ as a function of $\overline\psi$ using equation \refeq{eq-psiQE}. The vertical dashed line separates the two solutions of $\overline\psi[\overline{E}_*]$ which correspond to unvirialized cells with positive specific heat for $\overline\psi < 0.86$ and mostly virialized cells with negative specific heat for $\overline\psi > 0.86$. $\overline{T}_*[\overline\psi]$ goes to infinity as $\overline{\psi}$ goes to $0$.
Right plot: $\overline{W}_*$ as a function of $\overline\psi$ using equations \refeq{eq-WsQE} and \refeq{eq-psiQE}. The vertical dashed line at $\overline\psi = 2/3$ separates the two solutions of $\overline\psi[\overline{W}_*]$.
}
\label{fig-psiconv}
\end{center}
\end{figure*}

To numerically calculate the probability and normalization, we rewrite equation \refeq{eq-pE} in terms of $\overline\psi$ using the change of variables
\begin{equation}\label{eq-pp}
P(\overline{\psi}|N) d\overline{\psi} = P(\overline{E}_*|N)\left|\frac{d\overline{E}_*}{d\overline{T}_*}\frac{d\overline{T}_*}{d\overline{\psi}}\right| d\overline{\psi}
\end{equation}
so that the normalization integral becomes
\begin{equation}\label{eq-ppNorm}
P_{N,\mathrm{norm}} = \int_{\overline\psi_\mathrm{min}}^{\overline\psi_\mathrm{max}} P(\overline\psi|N)d\overline\psi.
\end{equation}
Equation \refeq{eq-psiQE} indicates that the quasi-equilibrium limits of $\overline\psi$ are within the range $0 \leq \overline\psi < 1$. Thus, writing the probability in terms of $\overline\psi$ transforms the normalization integral into a definite integral and simplifies its numerical evaluation. We illustrate the relationship between $\overline{T}_*$ and $\overline\psi$ in figure \ref{fig-psiconv}.

The probability in terms of $\overline\psi$ is given by
\begin{equation}\label{eq-ppProb}
P(\overline\psi_1 < \overline\psi < \overline\psi_2) = \frac{1}{P_{N,\mathrm{norm}}} \int_{\overline\psi_1}^{\overline\psi_2} P(\overline\psi|N)d\overline\psi
\end{equation}
from which the probability $P(\overline{T}_\stx{1} < \overline{T}_* < \overline{T}_\stx{2})$ follows as
\begin{equation}\label{eq-PTsdef}
P(\overline{T}_\stx{1} < \overline{T}_* < \overline{T}_\stx{2}) = P(\overline\psi[\overline{T}_\stx{1}] < \overline\psi < \overline\psi[\overline{T}_\stx{2}]).
\end{equation}
Here we use equation \refeq{eq-psiQE} to get $\overline\psi$ as a function of $\overline{T}_*$.

The probability that a cell has an energy of $\overline{W}_*$ or $\overline{E}_*$ is complicated by the fact that $\overline W_*[\overline\psi]$ and $\overline E_*[\overline\psi]$ are double-valued. This is because virialized systems have a negative specific heat, given by
\begin{equation}\label{eq-CV1}
C_V = \frac{1}{N}\left(\frac{\partial E}{\partial T}\right)_{N,V} = \frac{1}{N}\left(\frac{\partial \overline E_*}{\partial \overline T_*}\right)_{N,V}
\end{equation}
so that $\overline E_*[\overline T_*]$ is double valued. This has a transition point from positive to negative at $\overline{T}_* = 0.54$~\citep{2004ApJ...608..636L} which corresponds to $\overline\psi = 0.86$. The double-valued nature of $\overline W_*[\overline\psi]$ follows from the definition of $\overline W_* = \overline E_* - \overline T_*$ and has a minimum at $\overline\psi = 2/3$. To illustrate this, we plot $\overline W_*[\overline\psi]$ in figure \ref{fig-psiconv}. 

To calculate the probability for $\overline{W}_*$ or $\overline{E}_*$, we add the probability for both solutions to get
\begin{equation}\label{eq-PWsdef}
\begin{split}
&P(\overline{W}_\stx{1} < \overline{W}_* < \overline{W}_\stx{2}) = P(\overline\psi_-[\overline{W}_\stx{1}] < \overline\psi < \overline\psi_-[\overline{W}_\stx{2}])\\
&~~ + P(\overline\psi_+[\overline{W}_\stx{1}] < \overline\psi < \overline\psi_+[\overline{W}_\stx{2}])
\end{split}
\end{equation}
where $\overline\psi_-[\overline{W}_*]$ and $\overline\psi_+[\overline{W}_*]$ describe the conversion between $\overline{W}_*$ to $\overline\psi$ for different solutions. The probability for $\overline{E}_*$ is similarly defined.

\begin{figure*}[tb]
\begin{center}
\includegraphics[width=\floatwidth]{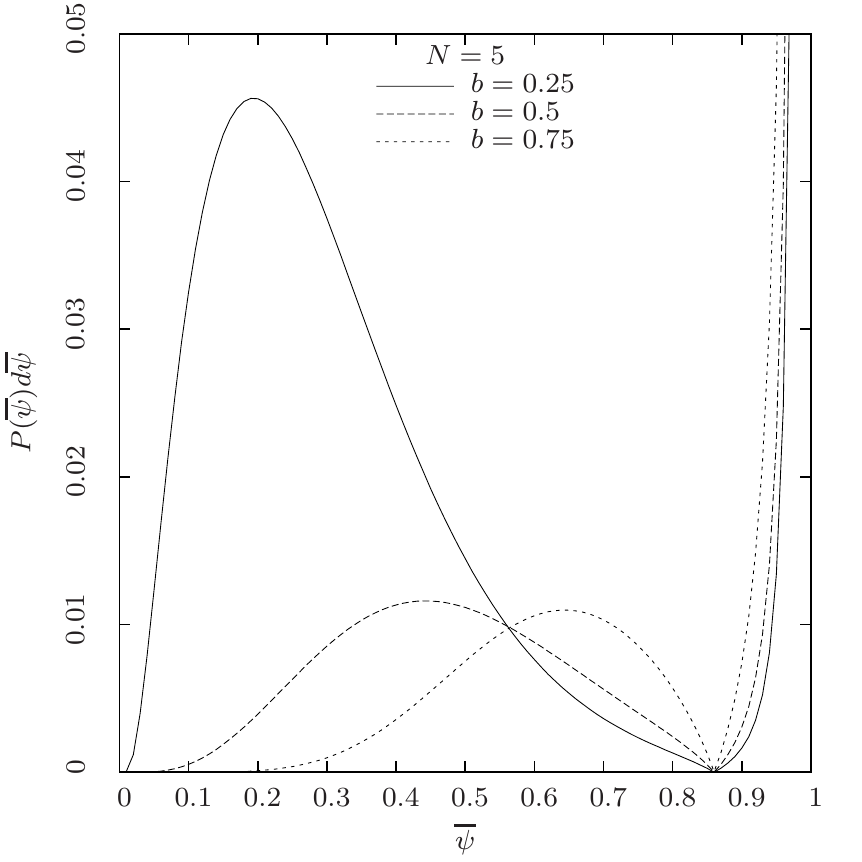}
\includegraphics[width=\floatwidth]{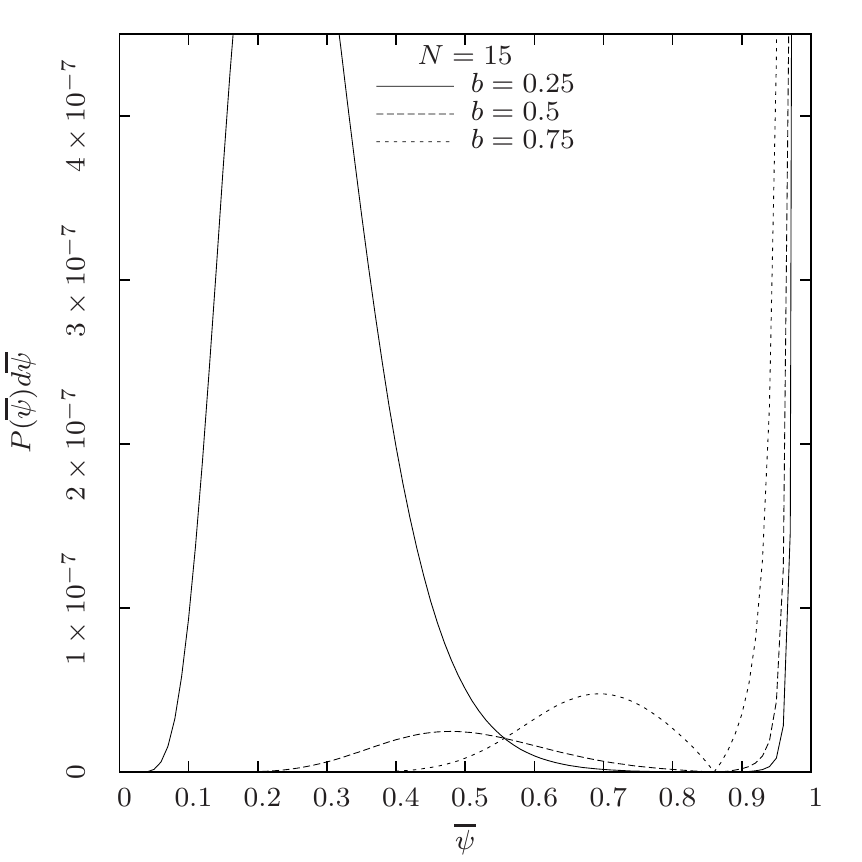}
\caption{
$P(\overline{\psi})$ for different values of $N$ and $b$ plotted as a function of $\overline{\psi}$. The left panel is for $N = 5$ and the right panel is for $N = 15$. The positive specific heat branch covers $\overline\psi < 0.86$ and the negative specific heat branch covers $\overline\psi > 0.86$.
}
\label{fig-psiprob}
\end{center}
\end{figure*}

These probabilities cover a wide range of conditions from virialized clusters to unbound collections of galaxies, and are described in detail in \citetalias{2012ApJ...745...87Y}. A key prediction of this theory is that most clusters are bound and virialized with negative specific heat, and clusters with more galaxies are very likely to have a negative specific heat. We illustrate these probabilities in figure \ref{fig-psiprob} for $N=5$ and $N=15$ and show that the negative specific heat branch dominates the probability, and the probability that a cell with $15$ galaxies has positive specific heat is negligible. Therefore we focus on cells with less than 20 galaxies because these cells have more pronounced features in the positive specific heat branch of the $P(\overline\psi)$ histogram.

\section{Catalog data}
\label{sec-data}

The New York University value-added galaxy catalog~(NYU-VAGC, \citealt{2005AJ....129.2562B}) is a composite catalog with the Sloan Digital Sky Survey~(SDSS) data as its primary component. It contains over 550,000 galaxies with their redshifts and positions on the sky. The catalog also contains extinction corrected and $K$-corrected absolute magnitudes for 8 bands, of which the $u$, $g$, $r$, $i$ and $z$ bands come from the SDSS and the $J$, $H$ and $K_s$ bands come from the 2-Micron All-Sky Survey~(2MASS) although for this study we use only the data from the SDSS. The galaxies in the catalog are also corrected for fibre collisions using the ``nearest'' method described in \citet{2005AJ....129.2562B}. Less than $10\%$ of the galaxies are affected by this correction which allows for a more complete sample in crowded regions.

In addition to the galaxy catalog, the NYU-VAGC also contains a survey geometry catalog that describes the survey footprint in terms of spherical polygons~(described in \citealt{2005AJ....129.2562B}). Since the SDSS is not an all-sky survey, the survey footprint determines the positions of cells and allows us to lay down cells where there is valid data.

For this work, we use the large scale structure samples in the version of the catalog corresponding to the seventh data release of the SDSS~\citep[DR7]{2009ApJS..182..543A}. We use the subsample with a flux limit of $r < 17.6$ and perform further selection cuts to obtain a volume and flux limited sample within a given redshift range.

To obtain a sufficiently large sample of galaxies, we use a low redshift sample in the range $0.01 \leq z \leq 0.12$. This redshift range contains a number of interesting structures including the Coma and Leo clusters, as well as the SDSS great wall. Using the criteria for completeness from \citet{2011ApJ...729..123Y}, the sample is complete in this redshift range for a $z$-band absolute magnitude $M_z < -20.8$. This limit is close to $L_*$ for a low redshift sample~\citep{2003ApJ...592..819B}.

Instead of the $r$-band data used in \citet{2011ApJ...729..123Y}, here we use the $z$-band because these galaxies are at a low redshift and the number counts in the $z$-band complete sample are comparable to the $r$-band sample. The other reason for using the $z$-band is that longer wavelengths are more sensitive to older stellar populations and are less prone to dust extinction. The $z$-band luminosities therefore provide a more consistent means of identifying galaxies of different morphologies and ages.

In addition to the complete sample, we use two other samples with a faint limit that is $1.2$ and $1.8$ magnitudes brighter. These brighter samples select for more massive galaxies that are more likely to dominate the potential of a cell and be the center of a large satellite system. If these satellite systems are virialized, they may be subclusters that form the building blocks of a rich cluster. This is particularly important because a comparison of bound and virialized probabilities in \citetalias{2012ApJ...745...87Y} suggests that clusters with more than about 10 members are very likely to be bound and virialized. These brighter galaxies may trace the positions of subclusters and allow us to represent a rich cluster by a small number of subcluster centers and thus obtain information about the larger scale assembly of clusters.

\begin{figure*}[tbp]
\begin{center}
\includegraphics[width=\floatwidth]{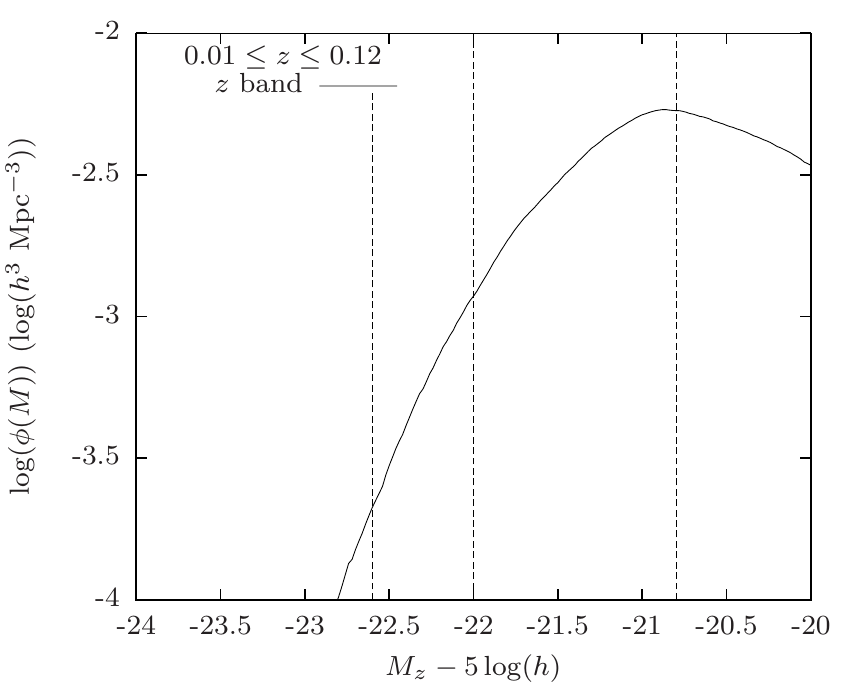}
\includegraphics[width=\floatwidth]{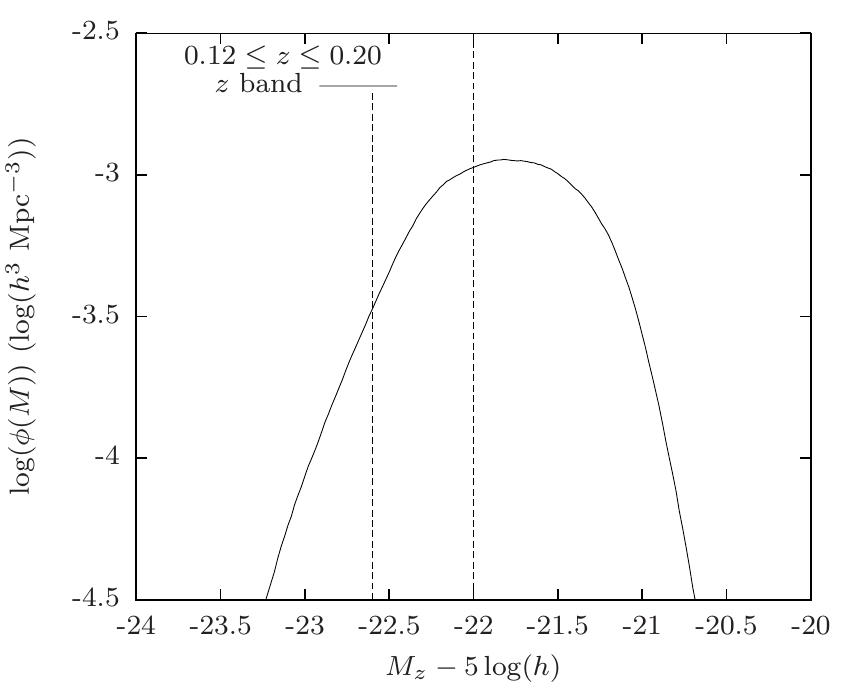}
\caption{
$z$-band Observed luminosity function for the NYU-VAGC at $0.01 \leq z \leq 0.12$~(left) and $0.12 \leq z \leq 0.20$~(right). The vertical lines indicate the absolute magnitude cuts we have adopted. Here $h=H_0/100$ km s$^{-1}$ Mpc$^{-1}$.
}
\label{fig-dataLF}
\end{center}
\end{figure*}

To increase the number of possible clusters for these brighter samples, we also consider an extended redshift range of $0.12 \leq z \leq 0.2$ for the brighter samples. We summarize these samples in table \ref{tab-gsample} and plot these limits on the observed luminosity function in figure \ref{fig-dataLF}.

\begin{deluxetable*}{l ccc c}
\tablewidth{0pt}
\tablecaption{\label{tab-gsample}
Selected subsamples
}
\tablehead{
\colhead{Sample}	& \colhead{Magnitude}		& \colhead{Redshift}	& \colhead{Density $\overline{n}$}	& \colhead{Galaxies} \\
	& \colhead{$M-5\log(h)$}	&	& \colhead{$h^{-3}$ Mpc$^3$}	&
}
\startdata
1a(z)	& $M_z < -20.8$	& $0.01 \leq z \leq 0.12$	& $4.56\times 10^{-3}$	& $149418$ \\
1b(z)	& $M_z < -22.0$	& $0.01 \leq z \leq 0.12$	& $4.13\times 10^{-4}$	& $13535$ \\
1c(z)	& $M_z < -22.6$	& $0.01 \leq z \leq 0.12$	& $5.34\times 10^{-5}$	& $1749$ \\
\\
2b(z)	& $M_z < -22.0$	& $0.12 \leq z \leq 0.20$	& $5.37\times 10^{-4}$	& $59162$ \\
2c(z)	& $M_z < -22.6$	& $0.12 \leq z \leq 0.20$	& $9.42\times 10^{-5}$	& $10388$
\enddata
\end{deluxetable*}

\section{Data Analysis and Procedure}
\label{sec-analysis}
Having selected subsamples of the NYU-VAGC for analysis, we identify clusters in the catalog, and determine the sample values of $\overline{N}$ and $b$. Here we define a cluster as a concentration of galaxies in a cell without regard to its virialization, and identify the cells that host these clusters. This definition of a cluster is particularly convenient because we can identify clusters solely on the basis of the positions of its members.

For our analysis, we select cell sizes that are representative of clusters in general. The cells are circular with a projected radius $R$ on the sky of 2.0, 5.0, 10.0 and 20.0 $h^{-1}$ Mpc where $h$ is related to the Hubble parameter $H_0$ by $h = H_0/100$. The smaller cell sizes describe the scale of a typical cluster or group of galaxies. The larger cell sizes describe the clustering of clusters and groups. Here, we describe such large structures as superclusters because they are essentially clusters of galaxy clusters and groups.

In the radial direction, the cells are defined by selecting velocity dispersions of $\Delta(cz)$ of 500, 1000 and 1500 km/s such that galaxies with a redshift within $\Delta(cz)$ of the cell's central redshift are considered members. This approximately cylindrical cell geometry reduces the effect of redshift space distortions by averaging over a range of redshifts and allows us to select cluster members by their peculiar velocities. Because of redshift space distortions, the use of spherical cells to identify clusters is not possible without a sufficiently precise secondary distance measure to all the galaxies in the cell.

To calculate $\overline{N}$ and $b$, we use the procedure in \citet{2011ApJ...729..123Y} with our essentially cylindrical cell geometry instead of spherical cells. Because we are interested in the average values of $\overline{N}$ and $b$, we do not calculate detailed error estimates for their inferred values, but note that because of cosmic variance, the estimated values of $\overline{N}$ and $b$ generally vary by about $25\%$ from quadrant to quadrant~\citep{2011ApJ...729..123Y}. We summarize our results in table \ref{tab-fvnstat_z}.

\begin{deluxetable}{lc rlc}
\tablewidth{0pt}
\tablecaption{\label{tab-fvnstat_z}
$z$-band Counts-in-cells $f_{V}(N)$
}
\tablehead{
\colhead{Sample}	&	\colhead{$\Delta (cz)$}	&	\colhead{Cells}	&	\colhead{$\overline{N}$}	&	\colhead{$b$}\\
	&	\colhead{km/s}	&	&	&	
}
\startdata
\hline
\multicolumn{5}{c}{$R = 2.0 h^{-1}$ Mpc} \\
\hline
1a(z)	&	\phn500	&	1157095	&	0.542	&	0.470	\\
1a(z)	&	1000	&	1107836	&	1.09	&	0.522	\\
1a(z)	&	1500	&	1059759	&	1.62	&	0.542	\\
1b(z)	&	\phn500	&	1157095	&	0.0486	&	0.154	\\
1b(z)	&	1000	&	1107836	&	0.0961	&	0.189	\\
1b(z)	&	1500	&	1059759	&	0.143	&	0.203	\\
1c(z)	&	\phn500	&	1157095	&	0.00625	&	0.037	\\
1c(z)	&	1000	&	1107836	&	0.0123	&	0.049	\\
1c(z)	&	1500	&	1059759	&	0.0182	&	0.055	\\
\hline
\multicolumn{5}{c}{$R = 5.0 h^{-1}$ Mpc} \\
\hline
1a(z)	&	\phn500	&	134329	&	\phn3.38	&	0.649	\\
1a(z)	&	1000	&	128923	&	\phn6.74	&	0.696	\\
1a(z)	&	1500	&	123334	&	10.1	&	0.712	\\
1b(z)	&	\phn500	&	134329	&	\phn0.302	&	0.306	\\
1b(z)	&	1000	&	128923	&	\phn0.597	&	0.360	\\
1b(z)	&	1500	&	123334	&	\phn0.884	&	0.381	\\
1c(z)	&	\phn500	&	134329	&	\phn0.0386	&	0.094	\\
1c(z)	&	1000	&	128923	&	\phn0.0760	&	0.121	\\
1c(z)	&	1500	&	123334	&	\phn0.111	&	0.135	\\
\hline
\multicolumn{5}{c}{$R = 10.0 h^{-1}$ Mpc} \\
\hline
1a(z)	&	\phn500	&	132662	&	13.4	&	0.741	\\
1a(z)	&	1000	&	132662	&	26.8	&	0.782	\\
1a(z)	&	1500	&	127476	&	40.1	&	0.796	\\
1b(z)	&	\phn500	&	132662	&	\phn1.18	&	0.421	\\
1b(z)	&	1000	&	132662	&	\phn2.36	&	0.491	\\
1b(z)	&	1500	&	127476	&	\phn3.50	&	0.517	\\
1c(z)	&	\phn500	&	132662	&	\phn0.147	&	0.154	\\
1c(z)	&	1000	&	132662	&	\phn0.295	&	0.209	\\
1c(z)	&	1500	&	127476	&	\phn0.436	&	0.234	\\
2b(z)	&	\phn500	&	133055	&	\phn1.57	&	0.460	\\
2b(z)	&	1000	&	133055	&	\phn3.15	&	0.526	\\
2b(z)	&	1500	&	127917	&	\phn4.76	&	0.551	\\
2c(z)	&	\phn500	&	133055	&	\phn0.271	&	0.222	\\
2c(z)	&	1000	&	133055	&	\phn0.541	&	0.280	\\
2c(z)	&	1500	&	127917	&	\phn0.811	&	0.302	\\
\hline
\multicolumn{5}{c}{$R = 20.0 h^{-1}$ Mpc} \\
\hline
1a(z)	&	\phn500	&	131562	&	\phn53.3	&	0.796	\\
1a(z)	&	1000	&	131562	&	106	&	0.834	\\
1a(z)	&	1500	&	131562	&	160	&	0.849	\\
1b(z)	&	\phn500	&	131562	&	\phn\phn4.61	&	0.507	\\
1b(z)	&	1000	&	131562	&	\phn\phn9.24	&	0.584	\\
1b(z)	&	1500	&	131562	&	\phn13.9	&	0.619	\\
1c(z)	&	\phn500	&	131562	&	\phn\phn0.566	&	0.221	\\
1c(z)	&	1000	&	131562	&	\phn\phn1.14	&	0.303	\\
1c(z)	&	1500	&	131562	&	\phn\phn1.72	&	0.343	\\
2b(z)	&	\phn500	&	132665	&	\phn\phn6.47	&	0.563	\\
2b(z)	&	1000	&	132665	&	\phn12.9	&	0.634	\\
2b(z)	&	1500	&	132665	&	\phn19.4	&	0.660	\\
2c(z)	&	\phn500	&	132665	&	\phn\phn1.09	&	0.312	\\
2c(z)	&	1000	&	132665	&	\phn\phn2.17	&	0.385	\\
2c(z)	&	1500	&	132665	&	\phn\phn3.25	&	0.415	\\
\enddata
\end{deluxetable}

\subsection{Cluster Identification}
To identify clusters and concentrations of galaxies, we use a modified version of the procedure described in section 2.2 of \citet{2009ApJS..183..197W} to take into account different cell sizes. This procedure identifies the cells having the most galaxies in the sample with the cells most likely to host a cluster. The algorithm we use is as follows:

\begin{enumerate}
\item For each galaxy in the sample, we assume that it is the central galaxy of a cluster and count the number of galaxies within a projected distance of $R$ and a redshift range of $\Delta (cz)$. We make no assumption as to whether these galaxies are cluster members or background galaxies. However, with an appropriate choice of $\Delta (cz)$, most of the identified galaxies are cluster members. In addition, we require that at least 95\% of the projected cell area is within the SDSS footprint. This will exclude cells close to the edge of the SDSS footprint that may have uncertain counts.

\item To avoid repeated identifications and overlapping cells, we remove overlapping cells using the following procedure:
\begin{enumerate}
\item Sort the list of cells by number of galaxies in descending order. If two cells have the same number of galaxies, the cell with the brighter galaxy is placed first.
\item Select the first cell in the sorted list.
\item Remove all other cells that overlap with the selected cell.
\item Select the next cell in the list and repeat steps (c) and (d) until the end of the list is reached.
\end{enumerate}
This procedure places emphasis on rich clusters by preferring clusters that have more members. In such clusters, the pairwise distance is smaller than in sparser clusters, and therefore the members are more likely to be bound to each other.

\item To get a list of cells that host a cluster in the sample, we remove cells with $N = 0$ or $N = 1$.
\end{enumerate}

Using this procedure, we identify clusters in the catalog using the given samples and cell sizes. In these samples, we take special note of cells with $N \leq 20$ because these cells are more likely to have a non-negligible chance of being unvirialized. We summarize the results of the cluster detection algorithm in table \ref{tab-clstat_z} with the number of clusters and the number of galaxies in the densest cell. The results show that the 1c(z) and 2c(z) samples indeed traces only the brightest galaxies with most 1c(z) and 2c(z) cells having less than 20 galaxies.

\begin{deluxetable}{lc rrc}
\tablewidth{0pt}
\tablecaption{\label{tab-clstat_z}
$z$-band Summary of Clusters
}
\tablehead{
\colhead{Sample}	&	\colhead{$\Delta (cz)$}	&	\colhead{Clusters}	&	\colhead{Clusters}	&	\colhead{Maximum $N$}\\
	&	\colhead{km/s}	&	\colhead{($N\leq20$)}	&	\colhead{(All)}	& 
}
\startdata
\hline
\multicolumn{5}{c}{$R = 2.0 h^{-1}$ Mpc} \\
\hline
1a(z)	& \phn500	& 16624	& 16760	& 47\\
1a(z)	& 1000		& 13654	& 13963	& 75\\
1a(z)	& 1500		& 11515	& 11901	& 92\\
1b(z)	& \phn500	& 2051	& 2051	& \phn9\\
1b(z)	& 1000		& 2132	& 2132	& 11\\
1b(z)	& 1500		& 2094	& 2094	& 13\\
1c(z)	& \phn500	& 125	& 125	& \phn4\\
1c(z)	& 1000		& 147	& 147	& \phn4\\
1c(z)	& 1500		& 161	& 161	& \phn5\\
\hline
\multicolumn{5}{c}{$R = 5.0 h^{-1}$ Mpc} \\
\hline
1a(z)	& \phn500	& 6122	& 7104	& 105\\
1a(z)	& 1000		& 3265	& 4586	& 170\\
1a(z)	& 1500		& 2046	& 3466	& 208\\
1b(z)	& \phn500	& 2260	& 2260	& \phn16\\
1b(z)	& 1000		& 2027	& 2029	& \phn21\\
1b(z)	& 1500		& 1802	& 1806	& \phn24\\
1c(z)	& \phn500	& 242	& 242	& \phn\phn6\\
1c(z)	& 1000		& 288	& 288	& \phn\phn7\\
1c(z)	& 1500		& 301	& 301	& \phn\phn7\\
\hline
\multicolumn{5}{c}{$R = 10.0 h^{-1}$ Mpc} \\
\hline
1a(z)	& \phn500	& 1165	& 2471	& 183\\
1a(z)	& 1000		& 271	& 1362	& 304\\
1a(z)	& 1500		& 103	& 963	& 367\\
1b(z)	& \phn500	& 1477	& 1489	& \phn29\\
1b(z)	& 1000		& 1020	& 1059	& \phn39\\
1b(z)	& 1500		& 780	& 835	& \phn45\\
1c(z)	& \phn500	& 295	& 295	& \phn\phn9\\
1c(z)	& 1000		& 318	& 318	& \phn13\\
1c(z)	& 1500		& 307	& 307	& \phn13\\
2b(z)	& \phn500	& 4893	& 4941	& \phn37\\
2b(z)	& 1000		& 3209	& 3398	& \phn54\\
2b(z)	& 1500		& 2316	& 2608	& \phn61\\
2c(z)	& \phn500	& 1654	& 1654	& \phn14\\
2c(z)	& 1000		& 1593	& 1593	& \phn15\\
2c(z)	& 1500		& 1455	& 1455	& \phn16\\
\hline
\multicolumn{5}{c}{$R = 20.0 h^{-1}$ Mpc} \\
\hline
1a(z)	& \phn500	& 31	& 733	& 413\\
1a(z)	& 1000		& 2	& 385	& 606\\
1a(z)	& 1500		& 0	& 260	& 758\\
1b(z)	& \phn500	& 593	& 656	& \phn58\\
1b(z)	& 1000		& 245	& 364	& \phn85\\
1b(z)	& 1500		& 129	& 262	& \phn95\\
1c(z)	& \phn500	& 303	& 303	& \phn14\\
1c(z)	& 1000		& 230	& 231	& \phn21\\
1c(z)	& 1500		& 191	& 192	& \phn23\\
2b(z)	& \phn500	& 1686	& 2008	& \phn68\\
2b(z)	& 1000		& 577	& 1122	& 103\\
2b(z)	& 1500		& 233	& 782	& 122\\
2c(z)	& \phn500	& 1257	& 1259	& \phn22\\
2c(z)	& 1000		& 887	& 892	& \phn25\\
2c(z)	& 1500		& 674	& 690	& \phn33\\
\enddata
\end{deluxetable}

\subsection{Cluster Energies}
To estimate the instantaneous scaled energies $T_*$ and $W_*$, we need further assumptions. The first is that the masses of all the galaxies in the cell are approximately equal to their average mass. This is generally a reasonable assumption for our purpose because the estimated values of $T_*$ and $W_*$ are not very sensitive to uncertainties in the galaxy masses. For example, figure 6 of \citetalias{2012ApJ...745...87Y} suggests that a difference in the mass of a factor of a few will lead to a worst case error in $W_*$ of about $10\%$. This is mainly because uncertainties from the transverse velocities and the anisotropy of a cluster's shape are likely to contribute more importantly to uncertainties in the energy estimates.

Using the equal mass assumption, we can calculate an approximate center-of-mass for the cluster in both projection and in redshift space. Using this center-of-mass, we determine a galaxy's position on the sky $r_\perp$ and radial peculiar velocity $v_\parallel$ with respect to the cluster's center-of-mass. This gives us structure information using the three observable quantities in phase space. The other three quantities, namely the radial position $r_\parallel$ and the transverse velocities giving $v_\perp$, will have to be estimated from further assumptions.

The next assumption is that the point-mass approximation is a good approximation for the potential of an individual galaxy. This is generally true for the cells we use because the half-mass radius of a galaxy is small compared to the cell radius. Under this approximation, $\epsilon/R$ is small and the $\zeta(\epsilon/R_1)$ and $\kappa(r/\epsilon)$ factors are essentially unity.

\subsubsection{Kinetic Energy}

To estimate the instantaneous scaled kinetic energy $T_*$ for a cell in the sample, we suppose that the velocity distribution and spatial distribution averaged over the entire cell is isotropic. This assumption gives
\begin{equation}\label{eq-v2iso}
\langle v^2 \rangle = \langle v_\parallel^2 \rangle + \langle v_\perp^2 \rangle
\end{equation}
for total peculiar velocity $v$, radial velocity $v_\parallel$ and transverse velocity $v_\perp$. Here the angle brackets denote the average over the cell. However, because the transverse velocity cannot be observed, we must use a free parameter, $\nu$, to describe the transverse velocity so that
\begin{equation}\label{eq-velsep}
\langle v^2 \rangle = \langle v^2_\parallel \rangle + \langle v^2_\perp \rangle = \langle v^2_\parallel \rangle\left(1+\frac{\langle v^2_\perp \rangle}{\langle v^2_\parallel \rangle}\right) \equiv \langle v^2_\parallel \rangle \nu^2.
\end{equation}
In the case of a cell with isotropic velocities, $\nu^2 = 3$.

To rescale the units of velocity to dimensionless units of cell radii per dynamical time, we estimate the dynamical time of the cell from its crossing time. A simple estimate of the crossing time is the time a test particle moving at a velocity $\sqrt{\langle v^2 \rangle}$ needs to traverse the radius of a cluster from its center of mass $\langle r \rangle$. Thus we have
\begin{equation}\label{eq-tcrossest}
\tau_\mathrm{dyn} \propto \tau_\mathrm{cross} \approx \frac{\sqrt{\langle r^2 \rangle}}{\sqrt{\langle v^2 \rangle}}
\end{equation}
where we define the dynamical time $\tau_\mathrm{dyn}$ as proportional to the estimated crossing time $\tau_\mathrm{cross}$. The isotropic assumption then gives
\begin{equation}\label{eq-tdyniso}
\tau_\mathrm{dyn} \approx k_\tau \sqrt{\frac{\langle r_\perp^2 \rangle}{2\langle v_\parallel^2 \rangle}}
\end{equation}
where $k_\tau$ is a free parameter that describes the relationship between the dynamical time and estimated crossing time and the uncertainties in the estimates of the cell crossing time. Using equation (71) of \citetalias{2012ApJ...745...87Y} for the scaled kinetic energy
\begin{equation}\label{eq-Tsups}
T_* = \frac{4}{9}\frac{1}{R^2}\frac{\langle (v\tau_\mathrm{dyn})^2 \rangle}{\zeta(\epsilon/R)},
\end{equation}
and equation \refeq{eq-tdyniso}, $T_*$ is
\begin{equation}\label{eq-TsObsdef}
T_* = \frac{4}{9}\upsilon^2 = \frac{4}{9}\nu^2k_\tau^2\frac{\tau_\mathrm{cross}^2}{R^2} \left\langle v_\parallel^2 \right\rangle
\end{equation}
where, $k_\tau$ and $\nu$ combine into a single dynamical free parameter.

\subsubsection{Correlation Potential Energy}
To estimate the correlation potential energy, we assume that it is essentially an extensive quantity. This is a reasonable approximation because at scales where the two-point correlation function is negligible, the expansion of the universe exactly cancels the smoothed background potential~\citep{1996ApJ...460...16S} so we can ignore the background contribution for cells that are larger than the scale at which the two-point correlation function $\xi_2$ is negligible. At smaller scales, \citet{1996ApJ...460...16S} show that extensivity is also a good approximation because the correlation energy within a cell is much greater than the correlation energy between cells. This means we can calculate the correlation potential energy by considering only gravitational interactions within the cell.

From \citetalias{2012ApJ...745...87Y}, the scaled correlation potential energy is~(c.f. equations (64), (65) of \citetalias{2012ApJ...745...87Y})
\begin{equation}\label{eq-WsObsdef1}
\begin{split}
W_* &= -\frac{8}{9N^2}\sum_{1\leq i < j \leq N} \left(\frac{1}{r_{ij}}\right)\\
 &= -\frac{8}{9N^2}\sum_{1\leq i < j \leq N} \left(\frac{R}{(r_{ij,\perp}^2 + r_{ij,\parallel}^2)^{1/2}}\right)
\end{split}
\end{equation}
where we separate the 3D pairwise separation $r_{ij} = (r_{ij,\perp}^2 + r_{ij,\parallel}^2)^{1/2}$ into its transverse component $r_{ij,\perp}$ and radial component $r_{ij,\parallel}$. Because we do not have precise distances to galaxies within a cell, we separate out the radial component with a parameter $\eta_{ij}$ such that
\begin{equation}\label{eq-possep}
r_{ij}^2 = r^2_{ij,\parallel} + r^2_{ij,\perp} = r^2_{ij,\perp} \left(1+\frac{r^2_{ij,\parallel}}{r^2_{ij,\perp}}\right) \equiv \eta_{ij}^2 r^2_{ij,\perp}.
\end{equation}
Substituting this into equation \refeq{eq-WsObsdef1} gives
\begin{equation}\label{eq-WsObsdef}
\begin{split}
W_* &= -\frac{8}{9N^2}\sum_{1\leq i < j \leq N} \left(\frac{R}{\eta_{ij}r_{ij,\perp}}\right) \\
&=  -\frac{4(N-1)}{9N}\frac{1}{\eta} \left\langle \frac{R}{r_{ij,\perp}} \right\rangle
\end{split}
\end{equation}
where $\eta$ is the averaged value of the individual $\eta_{ij}$ such that
\begin{equation}\label{eq-etadef}
\eta = \frac{\langle1/r_{ij}\rangle}{\langle1/r_{ij,\perp}\rangle}.
\end{equation}
For a uniform spherical cell, the average projected pairwise separation is approximately
\begin{equation} \label{eq-projunisep}
\begin{split}
R\left\langle \frac{1}{r_{ij,\perp}} \right\rangle &\approx \frac{R}{N \pi R^2}\int_0^R \frac{2 \pi r}{r} \sqrt{\frac{N}{\pi R^2}\left(R^2-r^2\right)}dr \\
&= \frac{1}{2}\sqrt{\frac{\pi}{N}}
\end{split}
\end{equation}
where $\langle 1/r_{ij,\perp} \rangle$ scales as $1/\sqrt{N}$ because denser and more clustered cells are more likely to have close pairs than less dense cells. Comparing this to the average pairwise separation $\langle 1/r_{ij} \rangle = 3/2$ for a spherical cell~\citet{2012ApJ...745...87Y}, we get $\eta \approx 1.69 \sqrt{N}$ for a uniform cell.

Although the cells we use are cylindrical, the clusters that the cells contain are likely to be spherical or elliptical. This means that these estimates are reasonable. Furthermore, actual clusters are also likely to have internal structures that will cause a departure from the uniform case. Therefore, the uniform and isotropic values of $\nu k_\tau$ and $\eta$ are a baseline for comparison, and may not necessarily represent a particular system.

\subsubsection{Correlation Virial Ratio}
From the kinetic energy and the correlation potential energy, the correlation virial ratio is
\begin{equation}\label{eq-psiObsdef1}
\psi = -\frac{W_*}{2T_*} = \frac{1}{\eta\nu^2k_\tau^2}\frac{(N-1)}{2N} \frac{ \left\langle R/r_{ij,\perp} \right\rangle }{(\tau_\mathrm{cross}^2/R^2) \left\langle v_\parallel^2 \right\rangle}
\end{equation}
which we write in terms of the observables $r_{ij,\perp}$ and $v_\parallel$ and the unobserved anisotropy and dynamical parameters $\eta$ and $\nu k_\tau$. Here again, we can group the anisotropy and dynamical parameters into a single modification factor $\eta\nu^2k_\tau^2$ and relate $\psi$ to the observables $N$, $R$, $r_\perp$ and $v_\parallel$. With independently determined values of $\eta$, $\nu$ and $k_\tau$, this relation provides a prediction for the observed virial ratio histogram.

\section{Comparisons with Theory}
\label{sec-comp}
To compare theory with observations, we calculate $T_*$, $W_*$ and $\psi$ for each cell containing an identified cluster using equations \refeq{eq-psiQE}, \refeq{eq-TsObsdef} and \refeq{eq-WsObsdef}. From these results, we can construct a histogram of scaled energies for cells with given $N$. To compare this histogram to observations, we compare the observed histograms to the quasi-equilibrium probabilities $P(\overline\psi_1 < \overline\psi < \overline\psi_2)$, $P(\overline{T}_\stx{1} < \overline{T}_* < \overline{T}_\stx{2})$ and $P(\overline{W}_\stx{1} < \overline{W}_* < \overline{W}_\stx{2})$.

However, the observations also contain significant uncertainties that will cause the estimated values of $T_*$ and $W_*$ to deviate from their true values. Some of these uncertainties, such as the anisotropy, come from inherent limitations in the observations. Others, such as departures of instantaneous values of $T_*$ and $W_*$ from their quasi-equilibrium values, are a result of dynamical fluctuations. Such fluctuations are an intrinsic property of a self gravitating system and early $N$-body simulations have shown that they may be as large as $20\%$~\citep{1972ApJ...172...17A}. Because of these fluctuations, it is reasonable to expect that the instantaneous properties of physical systems will only agree with average values on a statistical basis.

We can minimize the effect of some of these uncertainties which result from the free parameters $\nu k_\tau$ and $\eta$ which we expect to be of order unity. To do this, we determine the values of $\nu k_\tau$ and $\eta$ that minimize the total least squares distance between the expected and observed histograms. This provides an overall statistical correction to the anisotropy, but does not remove the intrinsic scatter that results from differences in the detailed structure of a cell and chance fluctuations in the phase space configurations of cells.

To incorporate these uncertainties using a simple model, we convolve the expected probabilities of $T_*$, $W_*$ and $\psi$ with normal distributions having zero mean, and variances $\sigma_{T*}^2$, $\sigma_{W}*^2$ and $\sigma_\psi^2$. This is generally reasonable since these uncertainties are likely to be combinations of a variety of different independent factors that smooth out the theoretical distributions and lower their peaks. We therefore obtain the expected observed histograms by convolving the theoretical ones with normal distributions of the observed uncertainties:
\begin{equation}\label{eq-PTobsConv}
P(T_\stx{obs}) \sim P(\overline{T}_*(\nu k_\tau)^2)\star \mathrm{Normal}(0,\sigma_{T*}^2),
\end{equation}
\begin{equation}\label{eq-PWobsConv}
P(W_\stx{obs}) \sim P(\overline{W}_*/\eta)\star \mathrm{Normal}(0,\sigma_{W*}^2)
\end{equation}
and
\begin{equation}\label{eq-PpobsConv}
P(\psi_\mathrm{obs}) \sim P(\overline\psi/(\eta\nu^2k_\tau^2))\star \mathrm{Normal}(0,\sigma_\psi^2).
\end{equation}
Because these normal distributions represent combinations of various observational uncertainties and dynamical fluctuations, the variances $\sigma_T*^2$, $\sigma_W*^2$ and $\sigma_\psi^2$ do not have simple interpretations. Therefore we focus our analysis on the anisotropy and dynamical parameters $\eta$ and $\nu k_\tau$.

We can estimate these observational variances by searching for their values which minimize the least-squares distance between the observed and theoretical histograms for $T_*$ and $W_*$. To ensure that there are enough clusters for reasonable statistics, we consider samples with at least 50 clusters and use histograms with 32 bins to smooth out fluctuations in the observations. Then, we use the inferred values of $\eta$ and $\nu k_\tau$ in equation \refeq{eq-PpobsConv} to determine the best-fit value of $\sigma_\psi$ and thus the expected histogram for $\psi$.

\begin{figure*}[tbp]
\begin{center}
\includegraphics[width=\floatwidth]{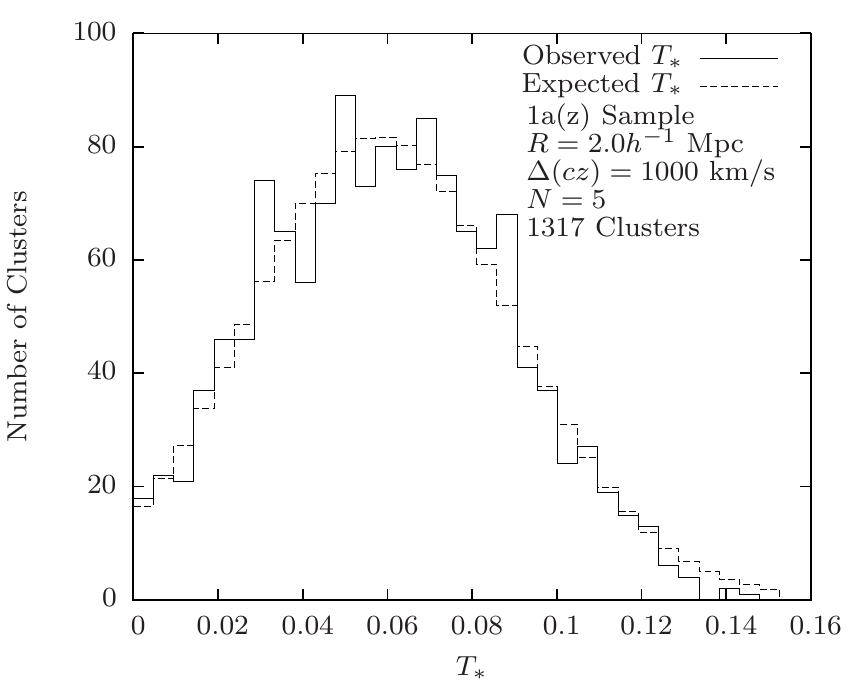}
\includegraphics[width=\floatwidth]{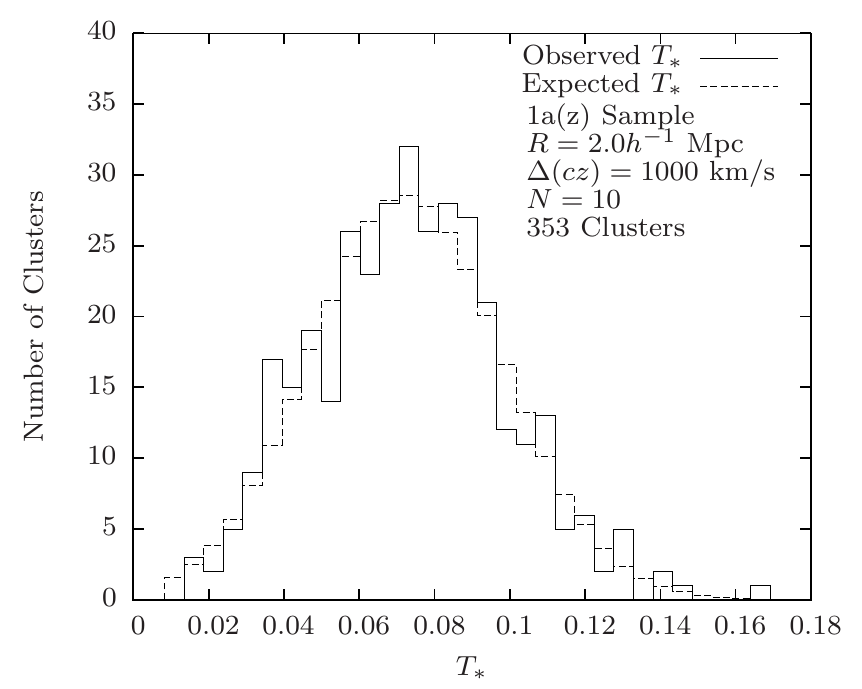}\\
\includegraphics[width=\floatwidth]{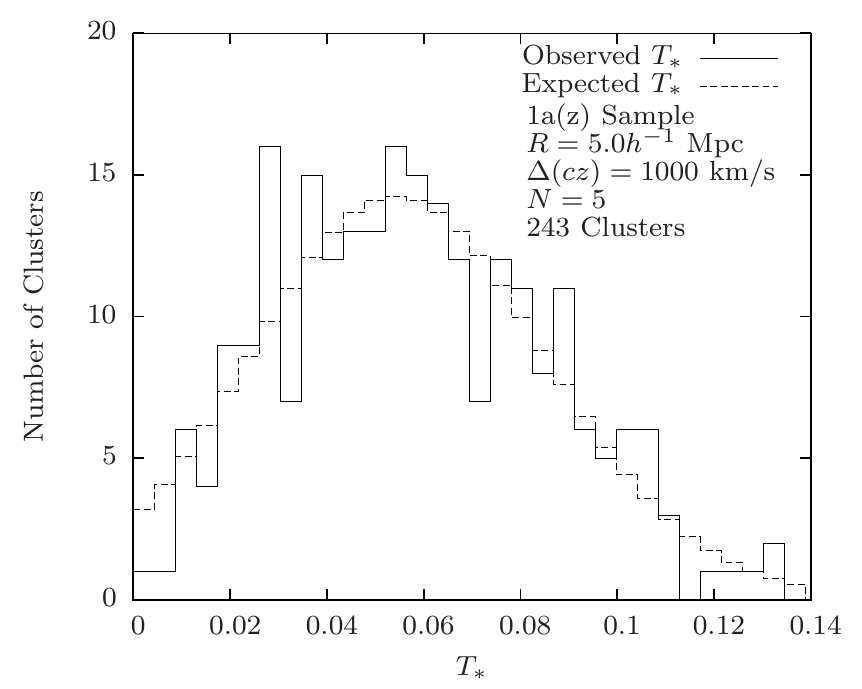}
\includegraphics[width=\floatwidth]{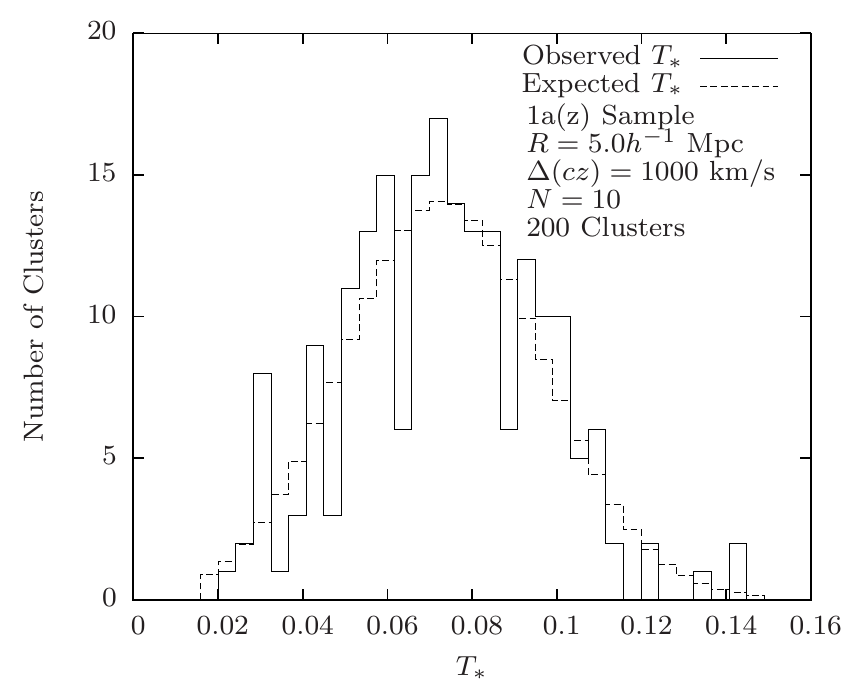}\\
\includegraphics[width=\floatwidth]{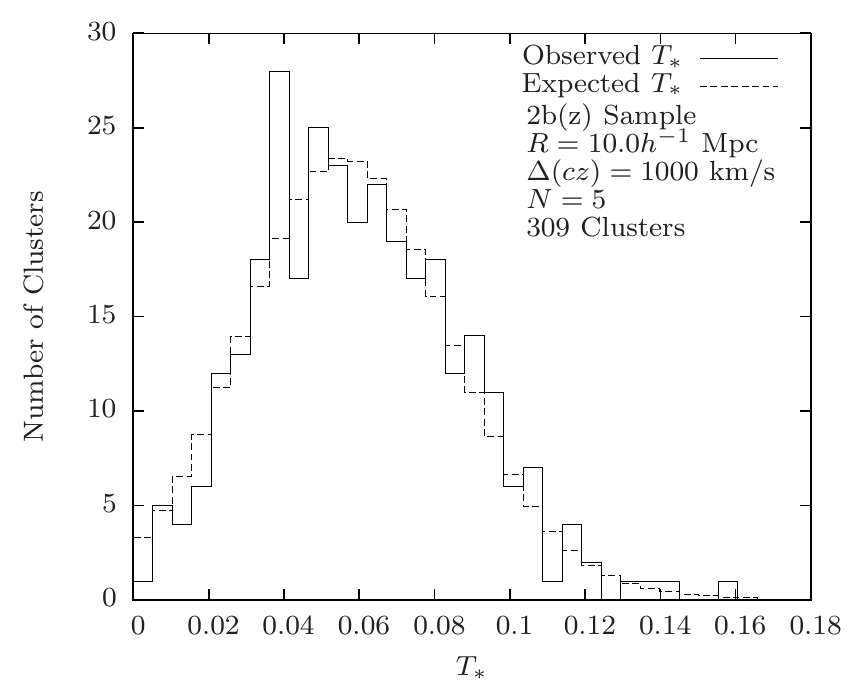}
\includegraphics[width=\floatwidth]{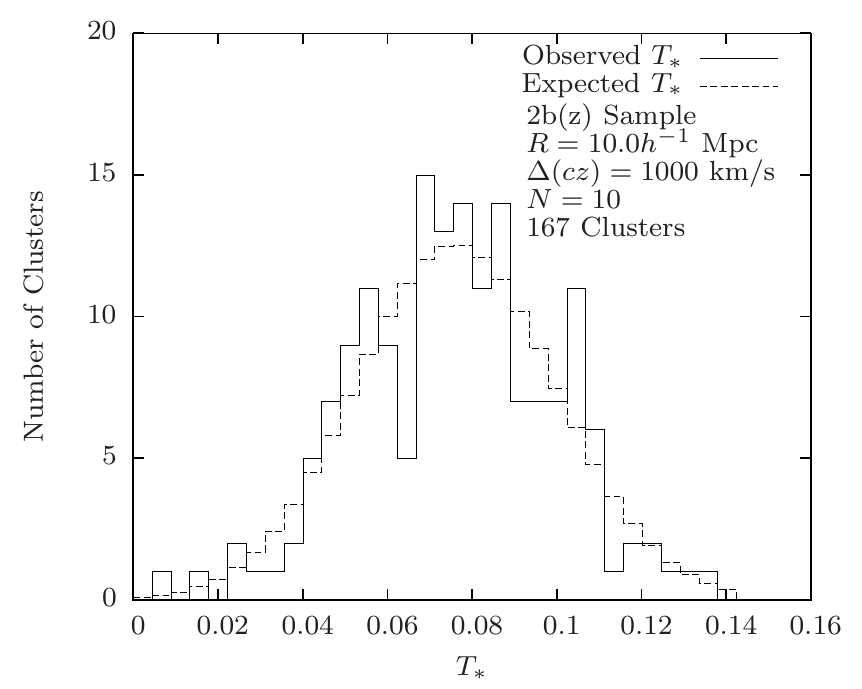}
\caption{
Observed and predicted histograms for $T_*$ for cells with 1000 km/s velocity dispersion. The solid line is the observed histogram and the dashed line is the expected histogram. Top row: 1a(z) sample, $R=2.0 h^{-1}$ Mpc; Middle row: 1a(z) sample, $R=5.0 h^{-1}$ Mpc; Bottom row: 2b(z) sample, $R=10.0 h^{-1}$ Mpc. Left column: $N=5$; Right column: $N=10$.
}
\label{fig-hist_Ts}
\end{center}
\end{figure*}

\begin{figure*}[tbp]
\begin{center}
\includegraphics[width=\floatwidth]{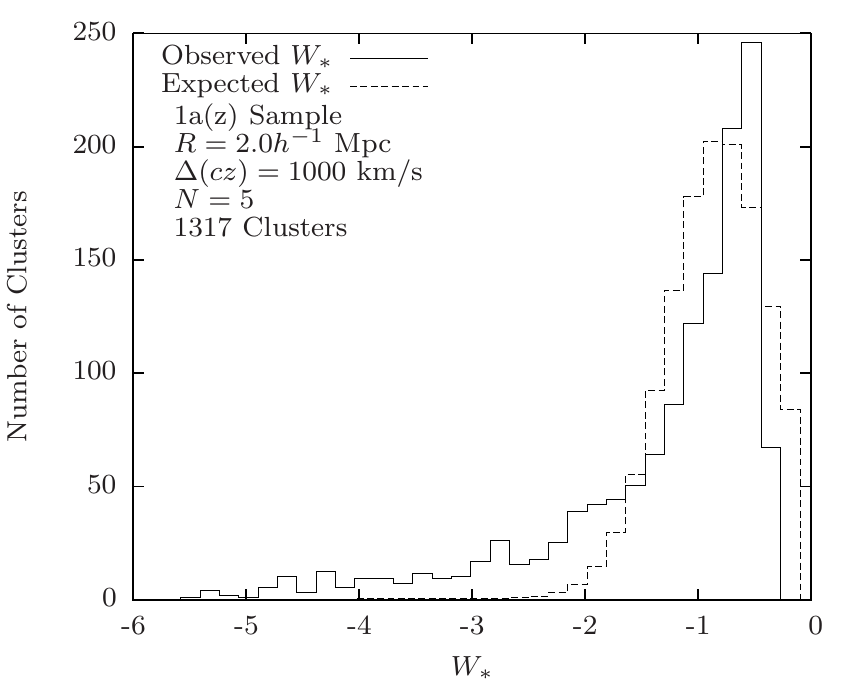}
\includegraphics[width=\floatwidth]{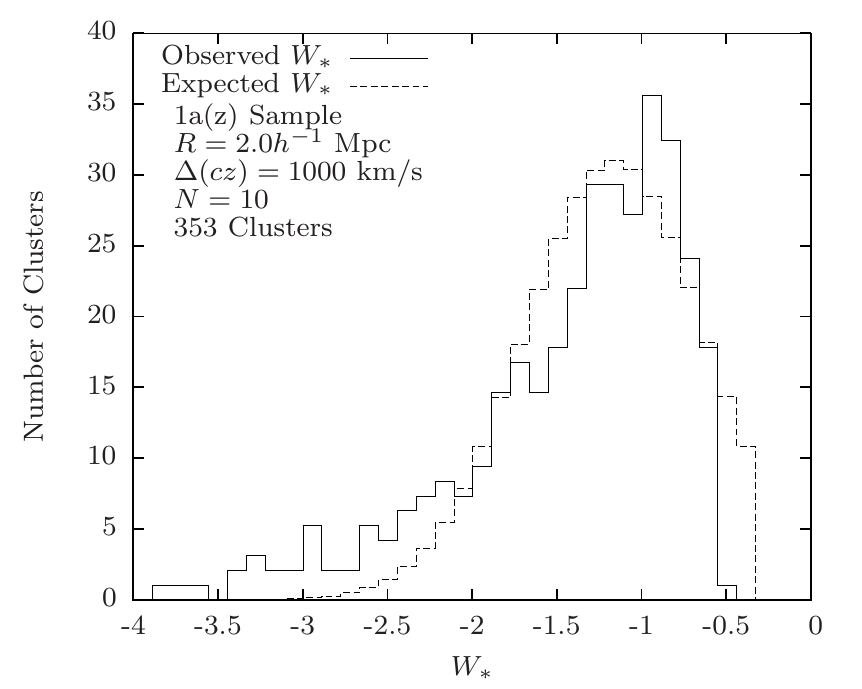}\\
\includegraphics[width=\floatwidth]{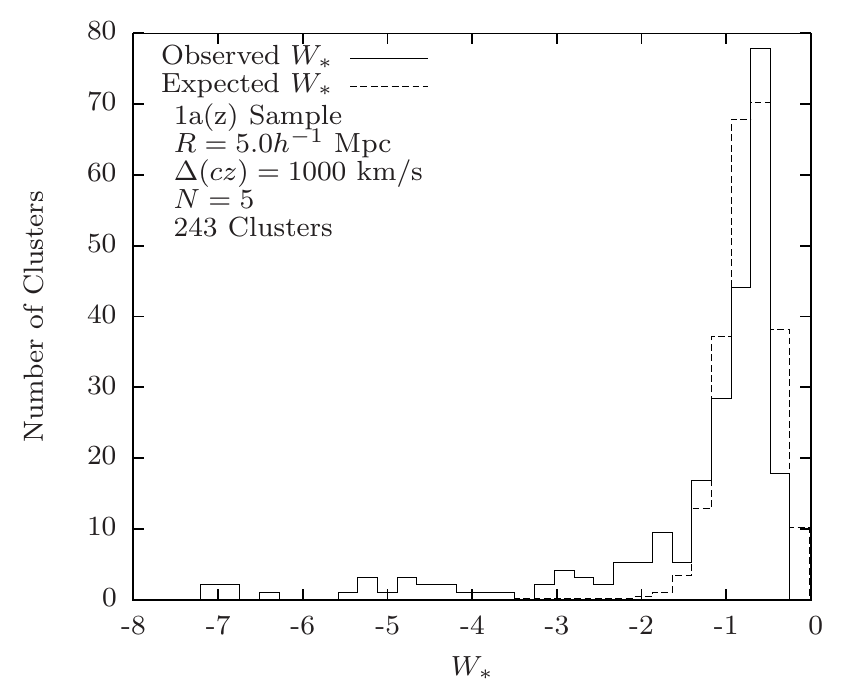}
\includegraphics[width=\floatwidth]{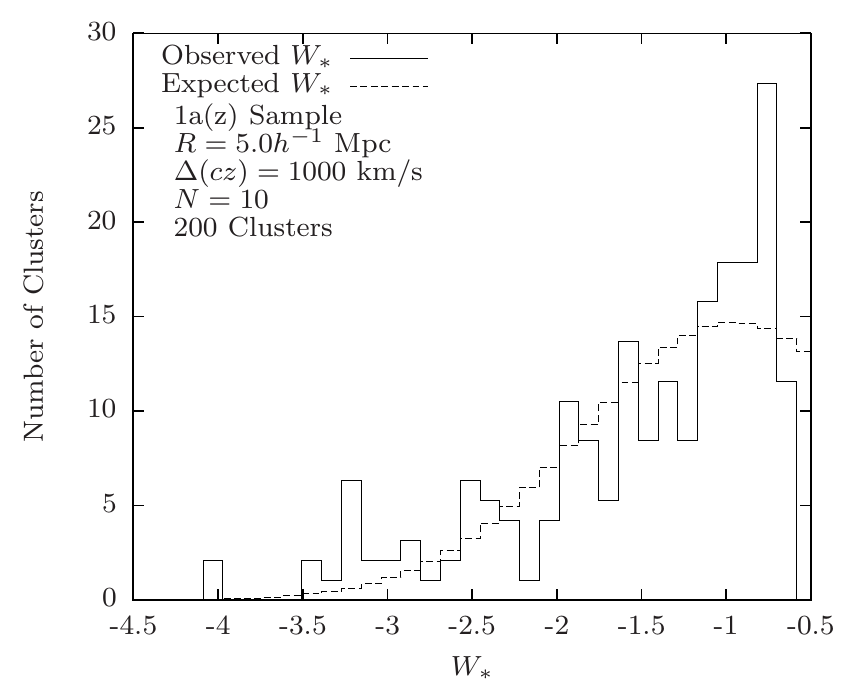}\\
\includegraphics[width=\floatwidth]{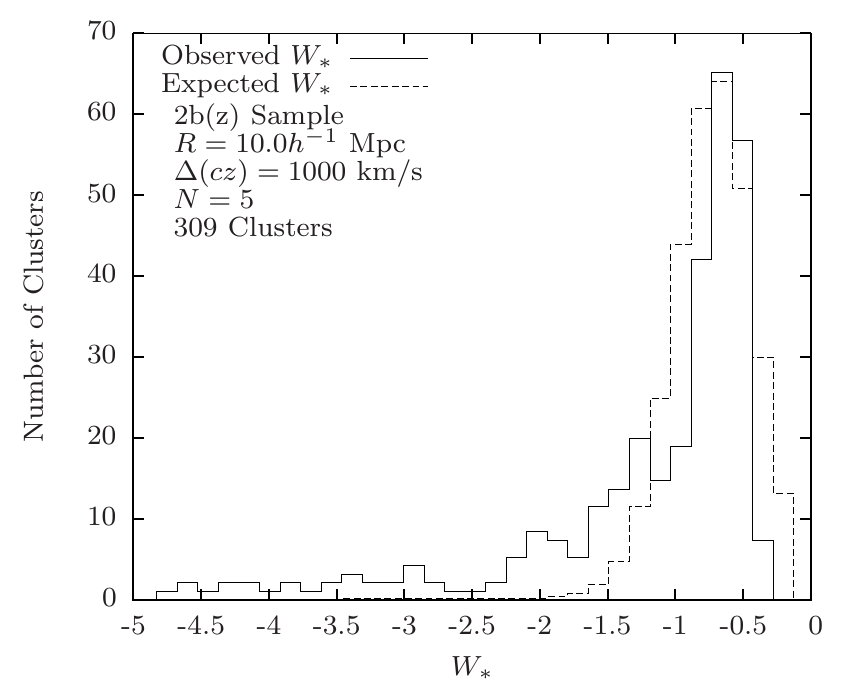}
\includegraphics[width=\floatwidth]{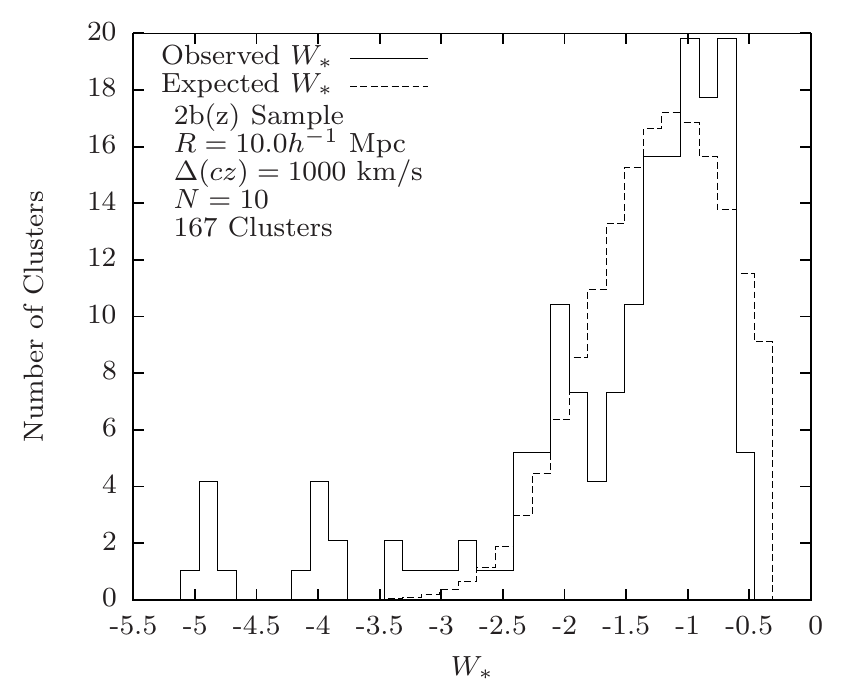}
\caption{
Observed and predicted histograms for $W_*$ for cells with 1000 km/s velocity dispersion. The solid line is the observed histogram and the dashed line is the expected histogram. Top row: 1a(z) sample, $R=2.0 h^{-1}$ Mpc; Middle row: 1a(z) sample, $R=5.0 h^{-1}$ Mpc; Bottom row: 2b(z) sample, $R=10.0 h^{-1}$ Mpc. Left column: $N=5$; Right column: $N=10$.
}
\label{fig-hist_Ws}
\end{center}
\end{figure*}

\begin{figure*}[tbp]
\begin{center}
\includegraphics[width=\floatwidth]{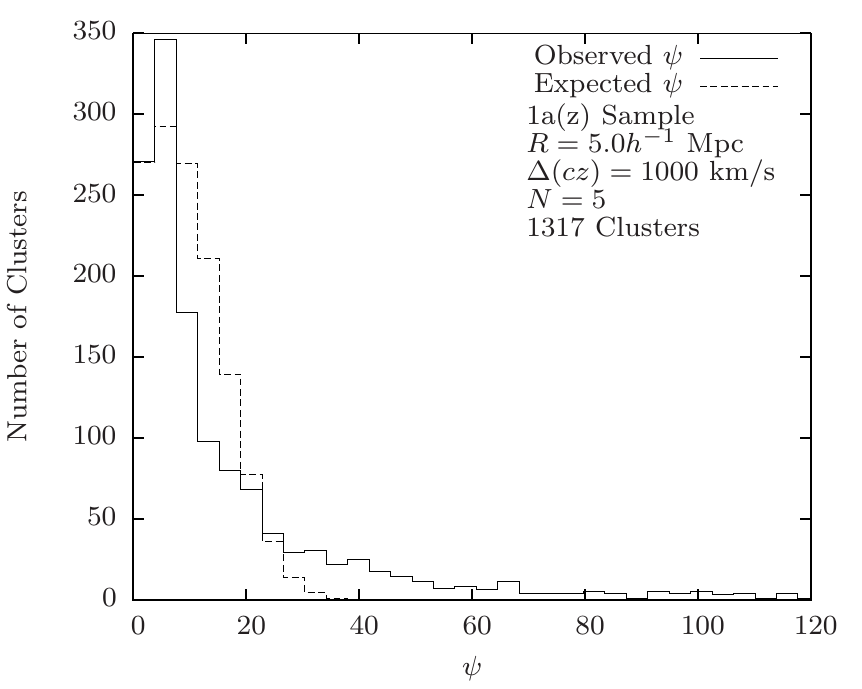}
\includegraphics[width=\floatwidth]{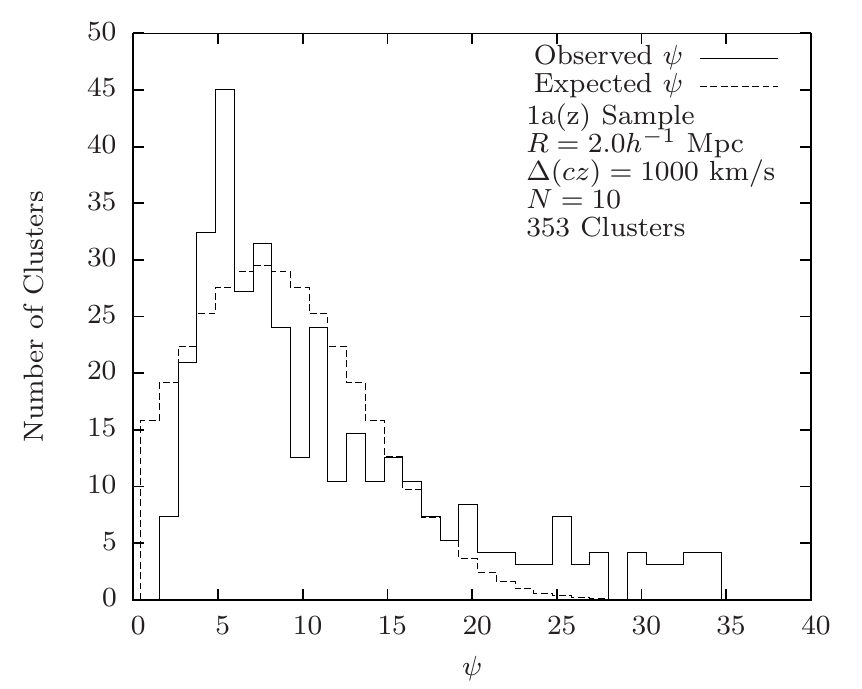}\\
\includegraphics[width=\floatwidth]{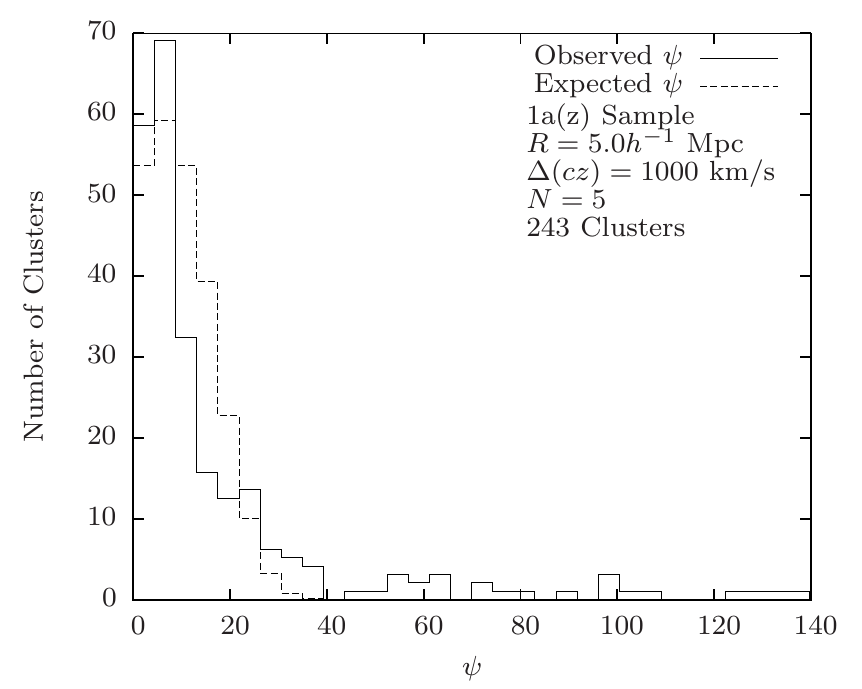}
\includegraphics[width=\floatwidth]{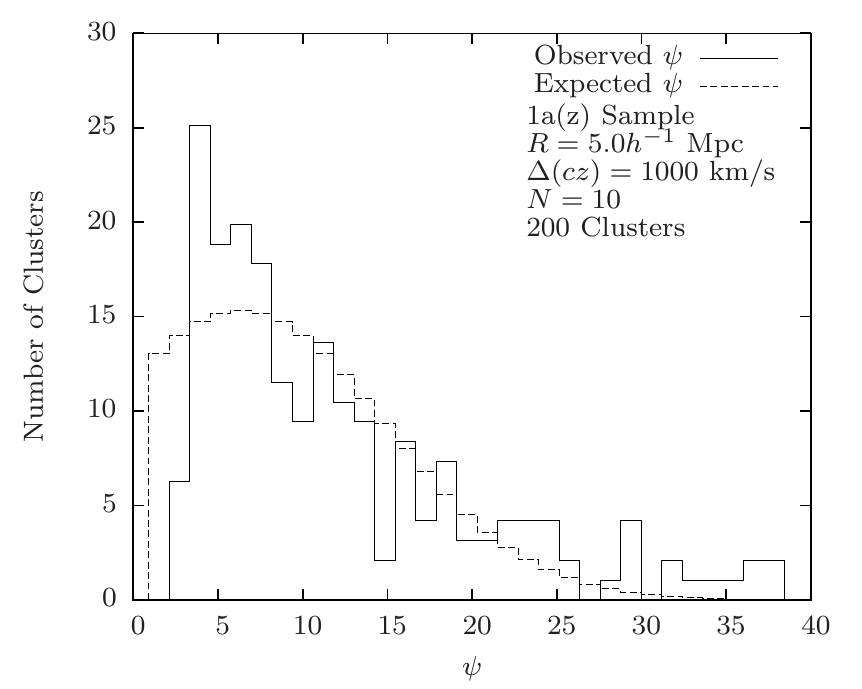}\\
\includegraphics[width=\floatwidth]{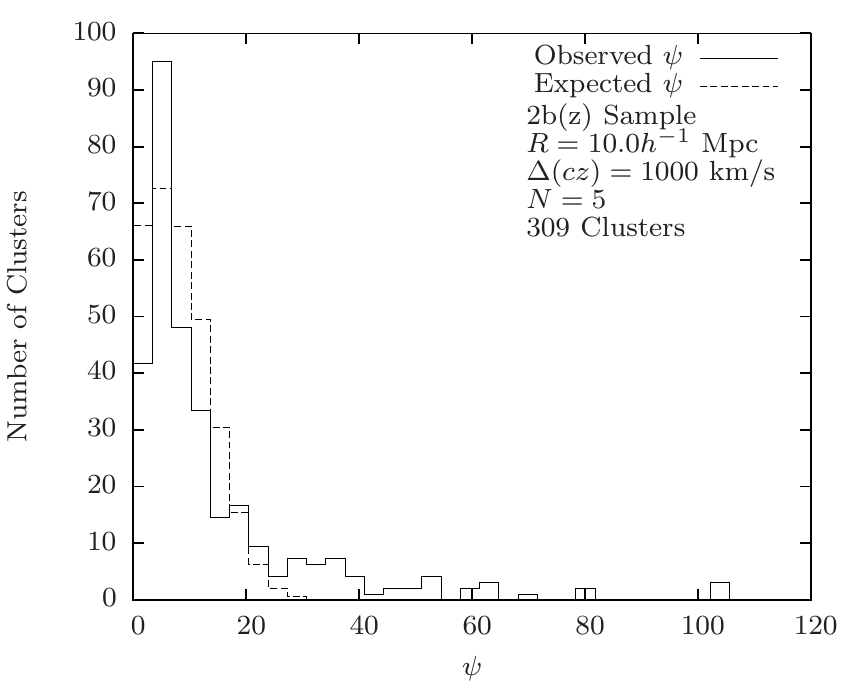}
\includegraphics[width=\floatwidth]{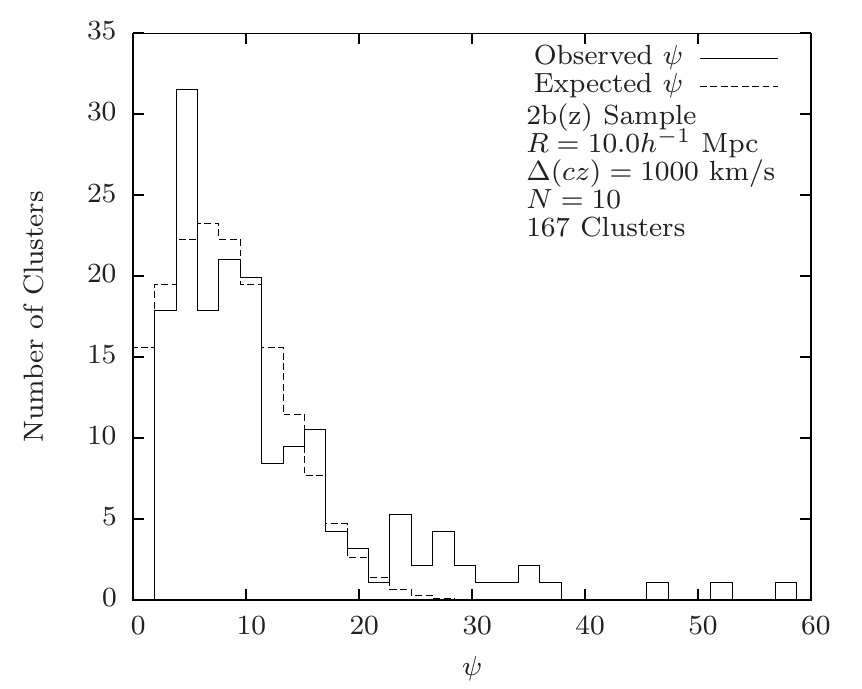}
\caption{
Observed and predicted histograms for $\psi$ for cells with 1000 km/s velocity dispersion. The solid line is the observed histogram and the dashed line is the expected histogram. Top row: 1a(z) sample, $R=2.0 h^{-1}$ Mpc; Middle row: 1a(z) sample, $R=5.0 h^{-1}$ Mpc; Bottom row: 2b(z) sample, $R=10.0 h^{-1}$ Mpc. Left column: $N=5$; Right column: $N=10$.
}
\label{fig-hist_ps}
\end{center}
\end{figure*}

To estimate uncertainties in the values of $\eta$, $\nu k_\tau$, $\sigma_{T*}^2$, $\sigma_{W*}^2$ and $\sigma_\psi^2$, we use a jackknife procedure that leaves out $10\%$ of the sample in each instance. For this analysis, we sort the clusters by galactic longitude and select a contiguous subsample that contains $90\%$ of the clusters. This resampling method incorporates spatial information and is a simple method to incorporate cosmic variance. Since there are less than 2000 clusters for each sample with given $N$, we calculate the resulting values for every possible subsample. This gives us an ensemble of values from which we can determine the 1-$\sigma$ range of each uncertain value. We summarize these results in tables \ref{tab-summary-N5}, \ref{tab-summary-N10} and \ref{tab-summary-N15} and plot some of these fits in figures \ref{fig-hist_Ts}, \ref{fig-hist_Ws} and \ref{fig-hist_ps}.

\begin{deluxetable*}{lc r ll ll ll}
\tablewidth{0pt}
\tabletypesize{\small}
\tablecaption{\label{tab-summary-N5}
Best fit $\eta$, $\nu k_\tau$, $\sigma_{W*}$, $\sigma_{T*}$ and $\sigma_\psi$, and uncertainties for $N=5$
}
\tablehead{
\colhead{Sample}	& \colhead{$\Delta (cz)$}	& \colhead{Clusters}	& \multicolumn{2}{c}{$W_*$ Histogram}	& \multicolumn{2}{c}{$T_*$ Histogram}	& \multicolumn{2}{c}{$\psi$ Histogram} \\
	& \colhead{km/s}	&	& \colhead{$\eta$}	& \colhead{$\sigma_W*$}	& \colhead{$\nu k_\tau$}	& \colhead{$\sigma_T*$}	& \colhead{$(\nu k_\tau)^2\eta$}	& \colhead{$\sigma_\psi$}
}
\startdata
\hline
\multicolumn{9}{c}{$R = 2.0 h^{-1}$ Mpc} \\
\hline
1a(z)	& \phn500	& 1614	& $3.74^{+0.35}_{-0.33}$	& $2.52^{+1.05}_{-0.64}$	& $1.47^{+0.071}_{-0.066}$	& $6.16^{+1.02}_{-1.00}$	& $8.08^{+1.64}_{-1.37}$	& $1.52^{+0.22}_{-0.20}$ \\
1a(z)	& 1000		& 1317	& $3.46^{+0.33}_{-0.42}$	& $2.56^{+1.18}_{-0.69}$	& $1.41^{+0.062}_{-0.060}$	& $6.31^{+0.92}_{-0.93}$	& $6.90^{+1.33}_{-1.35}$	& $2.18^{+0.66}_{-0.70}$ \\
1a(z)	& 1500		& 1204	& $3.15^{+0.32}_{-0.35}$	& $2.39^{+1.24}_{-0.69}$	& $1.39^{+0.056}_{-0.053}$	& $5.84^{+0.83}_{-0.85}$	& $6.04^{+1.17}_{-1.07}$	& $2.41^{+0.70}_{-0.78}$ \\
1b(z)	& \phn500	& 53	& $2.95^{+0.40}_{-0.54}$	& $4.29^{+2.29}_{-2.53}$	& $1.40^{+0.080}_{-0.063}$	& $9.55^{+2.84}_{-2.30}$	& $5.80^{+1.56}_{-1.48}$	& $2.53^{+1.65}_{-0.36}$ \\
1b(z)	& 1000		& 78	& $3.12^{+0.21}_{-0.14}$	& $1.79^{+0.61}_{-0.30}$	& $1.38^{+0.061}_{-0.060}$	& $6.55^{+1.01}_{-1.06}$	& $5.90^{+0.97}_{-0.75}$	& $1.45^{+0.49}_{-0.22}$ \\
1b(z)	& 1500		& 72	& $2.86^{+0.24}_{-0.20}$	& $1.11^{+0.64}_{-0.20}$	& $1.34^{+0.060}_{-0.056}$	& $6.59^{+1.12}_{-1.17}$	& $5.10^{+0.94}_{-0.75}$	& $2.10^{+0.82}_{-0.65}$ \\
\hline
\multicolumn{9}{c}{$R = 5.0 h^{-1}$ Mpc} \\
\hline
1a(z)	& \phn500	& 542	& $3.59^{+0.34}_{-0.40}$	& $1.33^{+0.62}_{-0.21}$	& $1.56^{+0.076}_{-0.079}$	& $6.13^{+1.02}_{-0.94}$	& $8.75^{+1.79}_{-1.75}$	& $1.68^{+0.09}_{-0.11}$ \\
1a(z)	& 1000		& 243	& $3.14^{+0.21}_{-0.15}$	& $1.11^{+0.33}_{-0.25}$	& $1.47^{+0.069}_{-0.065}$	& $6.73^{+1.08}_{-1.05}$	& $6.75^{+1.14}_{-0.89}$	& $1.67^{+0.53}_{-0.42}$ \\
1a(z)	& 1500		& 135	& $2.70^{+0.28}_{-0.16}$	& $1.91^{+0.57}_{-0.57}$	& $1.36^{+0.052}_{-0.056}$	& $6.36^{+0.97}_{-1.10}$	& $4.96^{+0.95}_{-0.67}$	& $1.59^{+0.69}_{-0.28}$ \\
1b(z)	& \phn500	& 186	& $3.30^{+0.27}_{-0.39}$	& $2.38^{+0.87}_{-0.55}$	& $1.30^{+0.049}_{-0.045}$	& $5.91^{+0.93}_{-0.91}$	& $5.60^{+0.93}_{-0.99}$	& $1.55^{+0.51}_{-0.45}$ \\
1b(z)	& 1000		& 209	& $2.96^{+0.37}_{-0.19}$	& $1.94^{+0.50}_{-0.60}$	& $1.25^{+0.041}_{-0.039}$	& $5.34^{+0.68}_{-0.74}$	& $4.58^{+0.91}_{-0.55}$	& $1.69^{+0.65}_{-0.29}$ \\
1b(z)	& 1500		& 202	& $2.81^{+0.27}_{-0.15}$	& $1.52^{+0.59}_{-0.46}$	& $1.26^{+0.039}_{-0.040}$	& $5.57^{+0.85}_{-0.89}$	& $4.44^{+0.74}_{-0.50}$	& $1.90^{+0.79}_{-0.39}$ \\
\hline
\multicolumn{9}{c}{$R = 10.0 h^{-1}$ Mpc} \\
\hline
1a(z)	& \phn500	& 76	& $3.92^{+0.41}_{-0.18}$	& $2.21^{+0.62}_{-0.42}$	& $1.82^{+0.135}_{-0.119}$	& $6.28^{+1.34}_{-1.37}$	& $13.0^{+3.57}_{-2.17}$	& $2.30^{+0.49}_{-0.48}$ \\
1b(z)	& \phn500	& 171	& $3.15^{+0.70}_{-0.13}$	& $1.21^{+0.50}_{-0.18}$	& $1.40^{+0.064}_{-0.062}$	& $6.92^{+1.12}_{-1.11}$	& $6.12^{+2.06}_{-0.76}$	& $1.49^{+0.44}_{-0.28}$ \\
1b(z)	& 1000		& 120	& $2.74^{+0.35}_{-0.21}$	& $1.96^{+0.92}_{-0.91}$	& $1.22^{+0.049}_{-0.047}$	& $6.66^{+0.84}_{-0.89}$	& $4.06^{+0.89}_{-0.60}$	& $1.34^{+1.09}_{-0.19}$ \\
1b(z)	& 1500		& 58	& $2.58^{+0.16}_{-0.10}$	& $2.29^{+1.15}_{-0.94}$	& $1.27^{+0.048}_{-0.050}$	& $8.11^{+1.73}_{-1.65}$	& $4.15^{+0.60}_{-0.48}$	& $2.15^{+1.23}_{-0.37}$ \\
2b(z)	& \phn500	& 516	& $3.67^{+0.22}_{-0.19}$	& $1.47^{+0.40}_{-0.36}$	& $1.58^{+0.075}_{-0.066}$	& $5.55^{+0.95}_{-0.91}$	& $9.14^{+1.50}_{-1.18}$	& $1.57^{+0.14}_{-0.16}$ \\
2b(z)	& 1000		& 309	& $3.04^{+0.35}_{-0.23}$	& $1.78^{+1.06}_{-0.47}$	& $1.43^{+0.058}_{-0.052}$	& $5.04^{+0.70}_{-0.71}$	& $6.24^{+1.29}_{-0.89}$	& $2.03^{+0.62}_{-0.68}$ \\
2b(z)	& 1500		& 177	& $2.81^{+0.19}_{-0.12}$	& $1.30^{+0.52}_{-0.25}$	& $1.40^{+0.061}_{-0.058}$	& $7.06^{+0.98}_{-0.96}$	& $5.50^{+0.88}_{-0.66}$	& $1.61^{+0.62}_{-0.43}$ \\
2c(z)	& \phn500	& 98	& $3.06^{+0.46}_{-0.14}$	& $1.63^{+0.89}_{-0.41}$	& $1.46^{+0.090}_{-0.086}$	& $8.12^{+1.54}_{-1.46}$	& $6.53^{+1.93}_{-1.01}$	& $2.35^{+0.77}_{-0.73}$ \\
2c(z)	& 1000		& 118	& $2.84^{+0.32}_{-0.15}$	& $1.90^{+0.41}_{-0.45}$	& $1.35^{+0.063}_{-0.057}$	& $5.47^{+0.98}_{-0.96}$	& $5.20^{+1.14}_{-0.67}$	& $0.91^{+0.14}_{-0.24}$ \\
2c(z)	& 1500		& 142	& $2.73^{+0.33}_{-0.20}$	& $1.93^{+0.59}_{-0.72}$	& $1.26^{+0.044}_{-0.049}$	& $4.97^{+0.69}_{-0.76}$	& $4.37^{+0.87}_{-0.62}$	& $2.49^{+0.71}_{-0.68}$ \\
\hline
\multicolumn{9}{c}{$R = 20.0 h^{-1}$ Mpc} \\
\hline
2b(z)	& \phn500	& 131	& $3.63^{+0.44}_{-0.42}$	& $1.16^{+0.26}_{-0.23}$	& $1.65^{+0.102}_{-0.081}$	& $6.66^{+1.72}_{-2.22}$	& $9.83^{+2.61}_{-1.98}$	& $1.71^{+0.42}_{-0.32}$ \\
2c(z)	& \phn500	& 136	& $2.99^{+0.68}_{-0.11}$	& $1.24^{+0.33}_{-0.24}$	& $1.32^{+0.071}_{-0.070}$	& $7.29^{+1.30}_{-1.29}$	& $5.20^{+1.89}_{-0.71}$	& $1.23^{+0.19}_{-0.18}$ \\
2c(z)	& 1000		& 98	& $3.32^{+0.20}_{-0.29}$	& $1.45^{+0.35}_{-0.30}$	& $1.29^{+0.060}_{-0.058}$	& $8.34^{+1.91}_{-1.55}$	& $5.49^{+0.89}_{-0.92}$	& $0.95^{+0.14}_{-0.15}$ \\
2c(z)	& 1500		& 65	& $2.68^{+0.45}_{-0.11}$	& $1.05^{+0.31}_{-0.27}$	& $1.25^{+0.040}_{-0.032}$	& $4.90^{+1.09}_{-0.90}$	& $4.18^{+1.02}_{-0.38}$	& $1.73^{+0.54}_{-0.57}$ \\
\enddata
\end{deluxetable*}

\begin{deluxetable*}{lc r ll ll ll}
\tablewidth{0pt}
\tabletypesize{\small}
\tablecaption{\label{tab-summary-N10}
Best fit $\eta$, $\nu k_\tau$, $\sigma_{W*}$, $\sigma_{T*}$ and $\sigma_\psi$, and uncertainties for $N=10$
}
\tablehead{
\colhead{Sample}	& \colhead{$\Delta (cz)$}	& \colhead{Clusters}	& \multicolumn{2}{c}{$W_*$ Histogram}	& \multicolumn{2}{c}{$T_*$ Histogram}	& \multicolumn{2}{c}{$\psi$ Histogram} \\
	& \colhead{km/s}	&	& \colhead{$\eta$}	& \colhead{$\sigma_W*$}	& \colhead{$\nu k_\tau$}	& \colhead{$\sigma_T*$}	& \colhead{$(\nu k_\tau)^2\eta$}	& \colhead{$\sigma_\psi$}
}
\startdata
\hline
\multicolumn{9}{c}{$R = 2.0 h^{-1}$ Mpc} \\
\hline
1a(z)	& \phn500	& 322	& $5.61^{+0.31}_{-0.34}$	& $4.12^{+1.02}_{-0.67}$	& $1.29^{+0.047}_{-0.041}$	& $5.95^{+0.68}_{-0.74}$	& $9.38^{+1.25}_{-1.12}$	& $4.04^{+1.13}_{-0.70}$ \\
1a(z)	& 1000		& 353	& $5.39^{+0.43}_{-0.44}$	& $4.74^{+1.17}_{-1.05}$	& $1.22^{+0.034}_{-0.029}$	& $4.82^{+0.63}_{-0.66}$	& $8.02^{+1.13}_{-0.99}$	& $5.37^{+1.84}_{-1.26}$ \\
1a(z)	& 1500		& 343	& $4.99^{+0.38}_{-0.35}$	& $5.10^{+1.33}_{-1.14}$	& $1.18^{+0.029}_{-0.034}$	& $4.77^{+0.56}_{-0.65}$	& $6.95^{+0.90}_{-0.85}$	& $5.00^{+1.44}_{-1.03}$ \\
\hline
\multicolumn{9}{c}{$R = 5.0 h^{-1}$ Mpc} \\
\hline
1a(z)	& \phn500	& 338	& $5.49^{+0.68}_{-0.71}$	& $4.83^{+1.63}_{-1.55}$	& $1.24^{+0.038}_{-0.036}$	& $5.56^{+0.67}_{-0.75}$	& $8.39^{+1.63}_{-1.50}$	& $3.71^{+1.36}_{-0.89}$ \\
1a(z)	& 1000		& 200	& $4.43^{+0.88}_{-1.62}$	& $7.65^{+2.68}_{-2.65}$	& $1.22^{+0.031}_{-0.030}$	& $5.52^{+0.72}_{-0.73}$	& $6.54^{+1.71}_{-2.60}$	& $7.04^{+1.49}_{-2.44}$ \\
1a(z)	& 1500		& 122	& $4.25^{+0.30}_{-0.51}$	& $3.02^{+2.34}_{-0.95}$	& $1.17^{+0.032}_{-0.030}$	& $6.43^{+0.88}_{-0.98}$	& $5.85^{+0.75}_{-0.96}$	& $3.74^{+1.27}_{-0.88}$ \\
\hline
\multicolumn{9}{c}{$R = 10.0 h^{-1}$ Mpc} \\
\hline
1a(z)	& \phn500	& 70	& $7.67^{+0.76}_{-1.10}$	& $3.02^{+0.75}_{-0.89}$	& $1.42^{+0.051}_{-0.047}$	& $5.96^{+0.96}_{-0.91}$	& $15.5^{+2.79}_{-3.09}$	& $1.96^{+0.20}_{-0.20}$ \\
1b(z)	& \phn500	& 55	& $4.46^{+0.41}_{-0.34}$	& $1.66^{+0.70}_{-0.30}$	& $1.17^{+0.032}_{-0.031}$	& $6.76^{+0.81}_{-0.87}$	& $6.08^{+0.93}_{-0.75}$	& $2.64^{+3.05}_{-0.45}$ \\
2b(z)	& \phn500	& 168	& $5.43^{+0.73}_{-0.31}$	& $3.45^{+0.87}_{-0.80}$	& $1.28^{+0.039}_{-0.040}$	& $6.31^{+0.81}_{-0.86}$	& $8.88^{+1.80}_{-1.03}$	& $2.39^{+0.64}_{-0.70}$ \\
2b(z)	& 1000		& 167	& $5.18^{+0.55}_{-0.43}$	& $4.27^{+1.23}_{-1.33}$	& $1.19^{+0.030}_{-0.030}$	& $5.15^{+0.64}_{-0.68}$	& $7.34^{+1.20}_{-0.94}$	& $3.31^{+1.00}_{-0.93}$ \\
2b(z)	& 1500		& 143	& $4.50^{+0.24}_{-0.45}$	& $2.46^{+0.92}_{-0.55}$	& $1.17^{+0.029}_{-0.029}$	& $5.38^{+0.70}_{-0.86}$	& $6.10^{+0.65}_{-0.89}$	& $3.06^{+0.75}_{-1.03}$ \\
\hline
\multicolumn{9}{c}{$R = 20.0 h^{-1}$ Mpc} \\
\hline
2b(z)	& \phn500	& 93	& $5.82^{+0.58}_{-0.60}$	& $1.46^{+0.30}_{-0.21}$	& $1.31^{+0.040}_{-0.041}$	& $5.51^{+0.85}_{-0.90}$	& $9.92^{+1.67}_{-1.57}$	& $2.40^{+1.51}_{-0.60}$ \\
\enddata
\end{deluxetable*}

\begin{deluxetable*}{lc r ll ll ll}
\tablewidth{0pt}
\tabletypesize{\small}
\tablecaption{\label{tab-summary-N15}
Best fit $\eta$, $\nu k_\tau$, $\sigma_{W*}$, $\sigma_{T*}$ and $\sigma_\psi$, and uncertainties for $N=15$
}
\tablehead{
\colhead{Sample}	& \colhead{$\Delta (cz)$}	& \colhead{Clusters}	& \multicolumn{2}{c}{$W_*$ Histogram}	& \multicolumn{2}{c}{$T_*$ Histogram}	& \multicolumn{2}{c}{$\psi$ Histogram} \\
	& \colhead{km/s}	&	& \colhead{$\eta$}	& \colhead{$\sigma_W*$}	& \colhead{$\nu k_\tau$}	& \colhead{$\sigma_T*$}	& \colhead{$(\nu k_\tau)^2\eta$}	& \colhead{$\sigma_\psi$}
}
\startdata
\hline
\multicolumn{9}{c}{$R = 2.0 h^{-1}$ Mpc} \\
\hline
1a(z)	& \phn500	& 94	& $6.08^{+0.37}_{-0.36}$	& $5.62^{+0.98}_{-0.82}$	& $1.20^{+0.033}_{-0.031}$	& $5.64^{+0.89}_{-0.92}$	& $8.81^{+1.05}_{-0.95}$	& $4.89^{+1.96}_{-0.89}$ \\
1a(z)	& 1000		& 120	& $5.74^{+0.27}_{-0.26}$	& $6.76^{+1.84}_{-1.49}$	& $1.21^{+0.031}_{-0.025}$	& $5.29^{+0.68}_{-0.69}$	& $8.38^{+0.85}_{-0.71}$	& $6.52^{+1.87}_{-1.34}$ \\
1a(z)	& 1500		& 119	& $5.39^{+0.35}_{-0.24}$	& $4.91^{+2.13}_{-0.93}$	& $1.17^{+0.026}_{-0.023}$	& $5.65^{+0.82}_{-0.82}$	& $7.32^{+0.82}_{-0.60}$	& $6.98^{+2.22}_{-1.88}$ \\
\hline
\multicolumn{9}{c}{$R = 5.0 h^{-1}$ Mpc} \\
\hline
1a(z)	& \phn500	& 174	& $6.94^{+0.51}_{-0.52}$	& $4.80^{+1.10}_{-1.09}$	& $1.20^{+0.031}_{-0.028}$	& $5.42^{+0.75}_{-0.79}$	& $10.1^{+1.30}_{-1.19}$	& $4.83^{+1.21}_{-0.80}$ \\
1a(z)	& 1000		& 140	& $6.03^{+0.37}_{-0.32}$	& $5.02^{+0.97}_{-0.76}$	& $1.11^{+0.013}_{-0.018}$	& $3.85^{+0.66}_{-0.70}$	& $7.46^{+0.65}_{-0.63}$	& $3.63^{+1.17}_{-0.61}$ \\
1a(z)	& 1500		& 117	& $5.68^{+0.59}_{-0.57}$	& $5.20^{+1.50}_{-1.16}$	& $1.09^{+0.021}_{-0.017}$	& $4.85^{+0.80}_{-0.77}$	& $6.76^{+0.98}_{-0.87}$	& $4.00^{+1.45}_{-0.87}$ \\
\hline
\multicolumn{9}{c}{$R = 10.0 h^{-1}$ Mpc} \\
\hline
1a(z)	& \phn500	& 63	& $6.91^{+0.50}_{-0.72}$	& $5.11^{+2.37}_{-1.16}$	& $1.25^{+0.024}_{-0.024}$	& $5.80^{+1.48}_{-1.35}$	& $10.8^{+1.23}_{-1.50}$	& $0.02^{+0.00}_{-0.00}$ \\
2b(z)	& \phn500	& 57	& $5.67^{+0.31}_{-0.31}$	& $2.49^{+0.77}_{-0.46}$	& $1.17^{+0.024}_{-0.026}$	& $6.46^{+1.21}_{-1.18}$	& $7.77^{+0.76}_{-0.76}$	& $2.56^{+1.30}_{-0.35}$ \\
2b(z)	& 1000		& 80	& $5.64^{+0.25}_{-0.27}$	& $7.02^{+1.85}_{-1.46}$	& $1.14^{+0.019}_{-0.019}$	& $5.92^{+0.82}_{-0.89}$	& $7.35^{+0.58}_{-0.58}$	& $6.44^{+1.93}_{-1.67}$ \\
2b(z)	& 1500		& 100	& $5.59^{+0.53}_{-0.41}$	& $5.53^{+1.04}_{-1.53}$	& $1.12^{+0.028}_{-0.030}$	& $5.85^{+0.64}_{-0.85}$	& $6.94^{+1.04}_{-0.85}$	& $4.88^{+2.09}_{-0.98}$ \\
\hline
\multicolumn{9}{c}{$R = 20.0 h^{-1}$ Mpc} \\
\hline
2b(z)	& \phn500	& 59	& $6.63^{+0.91}_{-0.24}$	& $3.14^{+1.16}_{-0.54}$	& $1.20^{+0.031}_{-0.033}$	& $5.74^{+0.77}_{-0.93}$	& $9.50^{+1.88}_{-0.84}$	& $4.73^{+1.65}_{-0.92}$ \\
\enddata
\end{deluxetable*}

\subsection{Discussion and Analysis}

The observed and expected histograms agree with each other in the case of $T_*$, but the observed $W_*$ and $\psi$ histograms have a long tail that is not accounted for by the theory. This long tail is a result of close galaxy pairs in $r_\perp$ that make a large contribution to the estimated $W_*$. These close pairs may be visual pairs that are separated by a large distance along the line-of-sight. Such pairs will have a large value of $\eta_{ij}$ that will increase the averaged value of $\eta$.

Other close pairs may be merging pairs that are better treated as a single extended galaxy since they are tightly bound and will eventually merge into a single galaxy~\citep{2010arXiv1011.0176Y}. For this reason, we can expect a population of outliers that have a very negative value of $W_*$. While it may be possible to identify merging pairs by their dynamics or morphologies, such an analysis is beyond the scope of this paper.

To quantitatively estimate the goodness of fit, we calculate a quantitative measure of the agreement between theory and observations using a one-sample Kolmogorov-Smirnov test~(e.g. \citealt{siegel1956nonparametric} chapter 4). We calculate the statistic $Z$ for a sample of $n$ clusters using
\begin{equation}\label{eq-Zstatdef}
Z = \sqrt{n}\sup_{i}|F_{\mathrm{obs},i}-F_{\mathrm{exp},i}|
\end{equation}
where $\sup_i$ denotes the largest absolute value of the difference between the cumulative observed and expected probabilities $F_{\mathrm{obs},i}$ and $F_{\mathrm{exp},i}$. These cumulative probabilities are defined as
\begin{equation}\label{eq-Fcumdef}
F_{\mathrm{obs},i} = \sum_{j\leq i} P_{\mathrm{obs},i}
\end{equation}
for observed probability $P_{\mathrm{obs},i}$ in bin $i$, and the expected cumulative probability $F_{\mathrm{exp},i}$ is similarly defined. To get the probability of obtaining a histogram that is as extreme as the observed histogram, or $p$-value, we compare the $Z$ statistic with the Kolmogorov-Smirnov distribution. We summarize the results of the Kolmogorov-Smirnov test for the $T_*$, $W_*$ and $\psi$ histograms in table \ref{tab-Tschi2}.

\begin{deluxetable*}{lcc r cr cr cr}
\tablewidth{0pt}
\tabletypesize{\small}
\tablecaption{\label{tab-Tschi2}
Kolmogorov-Smirnov test statistic and histogram probabilities
}
\tablehead{
\colhead{Sample}	& \colhead{$R$}	& \colhead{$\Delta (cz)$}	& \colhead{Clusters}	& \multicolumn{2}{c}{$W_*$ Histogram}	& \multicolumn{2}{c}{$T_*$ Histogram}	& \multicolumn{2}{c}{$\psi$ Histogram} \\
	& \colhead{$h^{-1}$ Mpc}	& km/s	&	& \colhead{$Z$}	& \colhead{$p$-value}	& \colhead{$Z$}	& \colhead{$p$-value}	& \colhead{$Z$}	& \colhead{$p$-value}
}
\startdata
\hline
\multicolumn{10}{c}{$N=5$} \\
\hline
1a(z)	& \phn2.0	& \phn500	& 1614	& 8.21	&  0.0\%	&  0.58	&  88.9\%	& 5.92	&  0.0\% \\
1a(z)	& \phn2.0	& 1000		& 1317	& 7.32	&  0.0\%	&  0.71	&  69.5\%	& 6.08	&  0.0\% \\
1a(z)	& \phn2.0	& 1500		& 1204	& 7.67	&  0.0\%	&  0.57	&  89.6\%	& 5.94	&  0.0\% \\
1b(z)	& \phn2.0	& \phn500	& 53	& 0.95	& 32.9\%	&  0.40	&  99.7\%	& 1.87	&  0.2\% \\
1b(z)	& \phn2.0	& 1000		& 78	& 2.19	&  0.0\%	&  0.37	&  99.9\%	& 2.35	&  0.0\% \\
1b(z)	& \phn2.0	& 1500		& 72	& 2.37	&  0.0\%	&  0.33	& 100.0\%	& 2.07	&  0.0\% \\[0.5em]
1a(z)	& \phn5.0	& \phn500	& 542	& 5.73	&  0.0\%	&  0.52	&  95.0\%	& 5.44	&  0.0\% \\
1a(z)	& \phn5.0	& 1000		& 243	& 3.56	&  0.0\%	&  0.42	&  99.5\%	& 2.57	&  0.0\% \\
1a(z)	& \phn5.0	& 1500		& 135	& 2.39	&  0.0\%	&  0.63	&  82.4\%	& 2.37	&  0.0\% \\
1b(z)	& \phn5.0	& \phn500	& 186	& 2.50	&  0.0\%	&  0.35	& 100.0\%	& 2.10	&  0.0\% \\
1b(z)	& \phn5.0	& 1000		& 209	& 2.52	&  0.0\%	&  0.46	&  98.2\%	& 3.13	&  0.0\% \\
1b(z)	& \phn5.0	& 1500		& 202	& 2.90	&  0.0\%	&  0.53	&  94.4\%	& 2.71	&  0.0\% \\[0.5em]
1a(z)	& 10.0	& \phn500	& 76	& 1.79	&  0.3\%	&  0.57	&  89.9\%	& 1.09	& 18.3\% \\
1b(z)	& 10.0	& \phn500	& 171	& 2.87	&  0.0\%	&  0.58	&  89.5\%	& 2.01	&  0.1\% \\
1b(z)	& 10.0	& 1000		& 120	& 2.62	&  0.0\%	&  0.43	&  99.3\%	& 3.18	&  0.0\% \\
1b(z)	& 10.0	& 1500		& 58	& 1.66	&  0.8\%	&  0.37	&  99.9\%	& 1.76	&  0.4\% \\
2b(z)	& 10.0	& \phn500	& 516	& 4.97	&  0.0\%	&  0.56	&  91.1\%	& 3.34	&  0.0\% \\
2b(z)	& 10.0	& 1000		& 309	& 4.71	&  0.0\%	&  0.42	&  99.4\%	& 2.95	&  0.0\% \\
2b(z)	& 10.0	& 1500		& 177	& 3.61	&  0.0\%	&  0.36	&  99.9\%	& 2.74	&  0.0\% \\
2c(z)	& 10.0	& \phn500	& 98	& 2.65	&  0.0\%	&  0.59	&  88.1\%	& 1.96	&  0.1\% \\
2c(z)	& 10.0	& 1000		& 118	& 1.78	&  0.3\%	&  0.40	&  99.7\%	& 1.49	&  2.3\% \\
2c(z)	& 10.0	& 1500		& 142	& 2.41	&  0.0\%	&  0.34	& 100.0\%	& 1.99	&  0.1\% \\[0.5em]
2b(z)	& 20.0	& \phn500	& 131	& 1.96	&  0.1\%	&  1.03	&  23.8\%	& 1.95	&  0.1\% \\
2c(z)	& 20.0	& \phn500	& 136	& 2.16	&  0.0\%	&  0.48	&  97.3\%	& 2.00	&  0.1\% \\
2c(z)	& 20.0	& 1000		& 98	& 1.88	&  0.2\%	&  0.46	&  98.4\%	& 1.89	&  0.2\% \\
2c(z)	& 20.0	& 1500		& 65	& 1.76	&  0.4\%	&  0.55	&  92.8\%	& 1.48	&  2.4\% \\
\hline
\multicolumn{10}{c}{$N=10$} \\
\hline
1a(z)	& \phn2.0	& \phn500	& 322	& 2.72	&  0.0\%	&  0.72	&  68.2\%	& 2.62	&  0.0\% \\
1a(z)	& \phn2.0	& 1000		& 353	& 2.05	&  0.0\%	&  0.29	& 100.0\%	& 2.52	&  0.0\% \\
1a(z)	& \phn2.0	& 1500		& 343	& 2.05	&  0.0\%	&  0.48	&  97.5\%	& 2.42	&  0.0\% \\[0.5em]
1a(z)	& \phn5.0	& \phn500	& 338	& 2.39	&  0.0\%	&  0.44	&  99.0\%	& 3.26	&  0.0\% \\
1a(z)	& \phn5.0	& 1000		& 200	& 1.14	& 14.7\%	&  0.33	& 100.0\%	& 1.47	&  2.7\% \\
1a(z)	& \phn5.0	& 1500		& 122	& 2.28	&  0.0\%	&  0.23	& 100.0\%	& 2.09	&  0.0\% \\[0.5em]
1a(z)	& 10.0	& \phn500	& 70	& 1.25	&  8.6\%	&  0.55	&  92.2\%	& 1.31	&  6.3\% \\
1b(z)	& 10.0	& \phn500	& 55	& 2.42	&  0.0\%	&  0.38	&  99.9\%	& 2.26	&  0.0\% \\
2b(z)	& 10.0	& \phn500	& 168	& 1.64	&  0.9\%	&  0.49	&  97.2\%	& 1.80	&  0.3\% \\
2b(z)	& 10.0	& 1000		& 167	& 1.98	&  0.1\%	&  0.29	& 100.0\%	& 1.82	&  0.3\% \\
2b(z)	& 10.0	& 1500		& 143	& 2.45	&  0.0\%	&  0.52	&  94.7\%	& 1.58	&  1.4\% \\[0.5em]
2b(z)	& 20.0	& \phn500	& 93	& 1.74	&  0.5\%	&  0.36	& 100.0\%	& 2.33	&  0.0\% \\
\hline
\multicolumn{10}{c}{$N=15$} \\
\hline
1a(z)	& \phn2.0	& \phn500	& 94	& 0.64	& 80.0\%	&  0.35	& 100.0\%	& 1.60	&  1.2\% \\
1a(z)	& \phn2.0	& 1000		& 120	& 1.12	& 16.5\%	&  0.61	&  85.5\%	& 1.10	& 17.5\% \\
1a(z)	& \phn2.0	& 1500		& 119	& 1.95	&  0.1\%	&  0.43	&  99.2\%	& 1.24	&  9.3\% \\[0.5em]
1a(z)	& \phn5.0	& \phn500	& 174	& 1.81	&  0.3\%	&  0.49	&  97.0\%	& 1.67	&  0.8\% \\
1a(z)	& \phn5.0	& 1000		& 140	& 1.24	&  9.3\%	&  0.59	&  87.7\%	& 2.31	&  0.0\% \\
1a(z)	& \phn5.0	& 1500		& 117	& 1.25	&  8.8\%	&  0.56	&  91.2\%	& 1.80	&  0.3\% \\[0.5em]
1a(z)	& 10.0	& \phn500	& 63	& 1.36	&  5.0\%	&  0.66	&  78.1\%	& 3.31	&  0.0\% \\
2b(z)	& 10.0	& \phn500	& 57	& 1.64	&  0.9\%	&  0.45	&  98.6\%	& 2.07	&  0.0\% \\
2b(z)	& 10.0	& 1000		& 80	& 0.69	& 72.3\%	&  0.36	& 100.0\%	& 1.04	& 23.2\% \\
2b(z)	& 10.0	& 1500		& 100	& 0.97	& 30.3\%	&  0.69	&  73.3\%	& 1.55	&  1.6\% \\[0.5em]
2b(z)	& 20.0	& \phn500	& 59	& 1.46	&  2.9\%	&  0.54	&  93.4\%	& 1.43	&  3.3\% \\
\enddata
\end{deluxetable*}

We find that we cannot reject the null hypothesis that $T_*$ follows the theoretical expected distribution for almost all the instances at the $95\%$ level. However, this is not the case for the $W_*$ and $\psi$ histograms where most instances are unlikely to agree with the theory. This is expected because of the abnormally long tails that come from the close two-dimensional projected galaxy pairs in a dense cluster.

\subsection{Comparison between cell sizes}
\begin{figure*}[tbp]
\begin{center}
\includegraphics[width=\floatwidth]{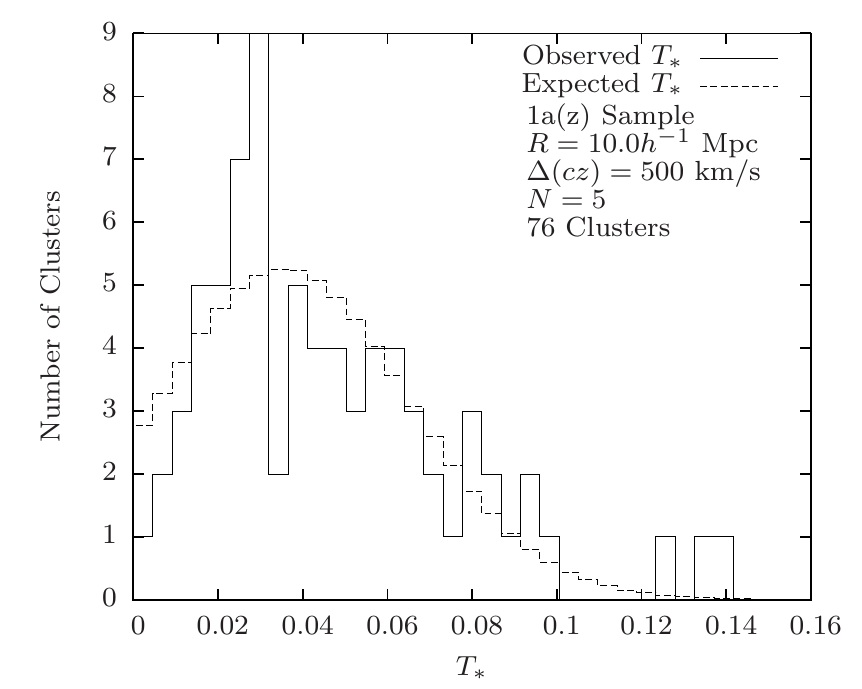}
\includegraphics[width=\floatwidth]{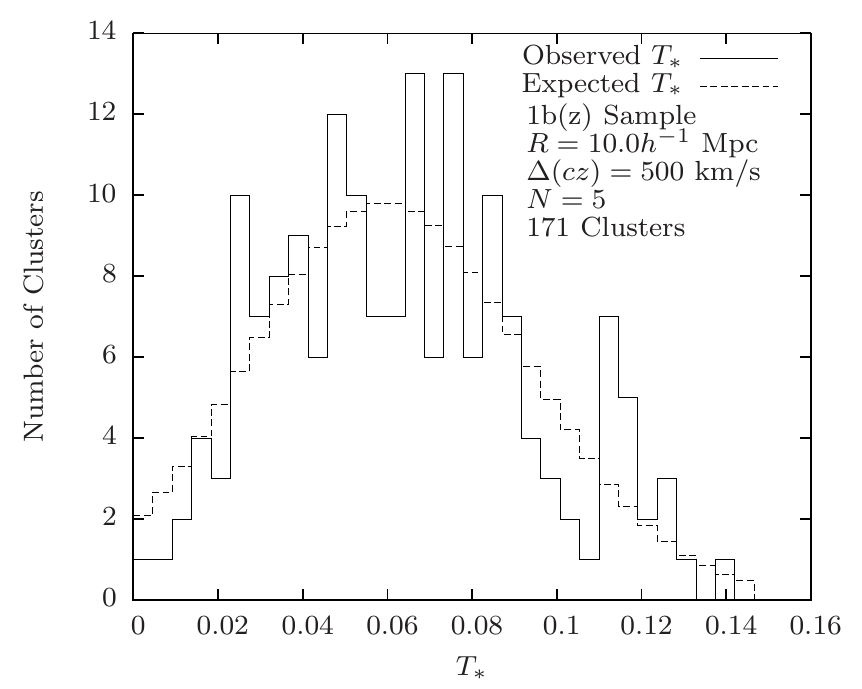}\\
\includegraphics[width=\floatwidth]{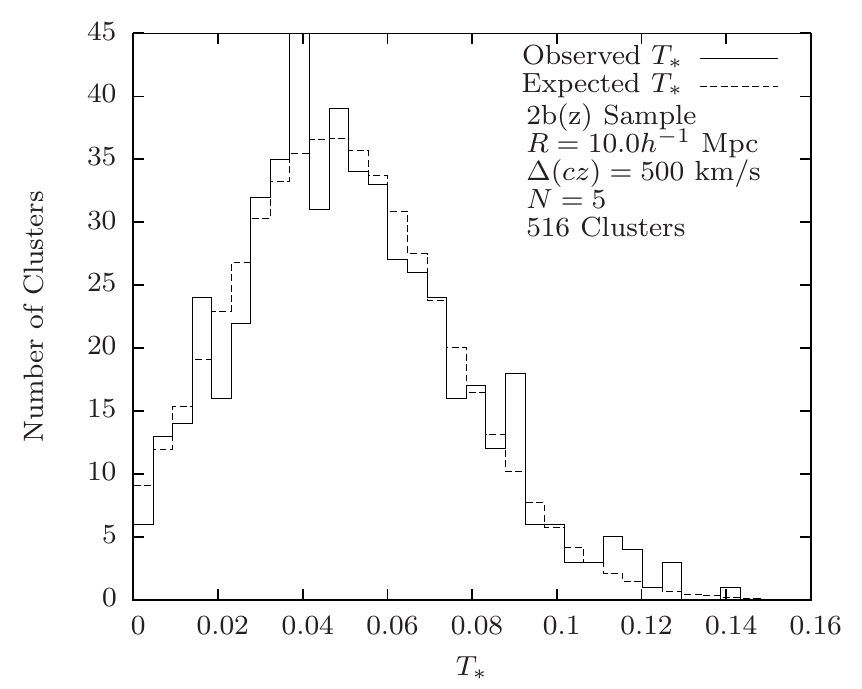}
\includegraphics[width=\floatwidth]{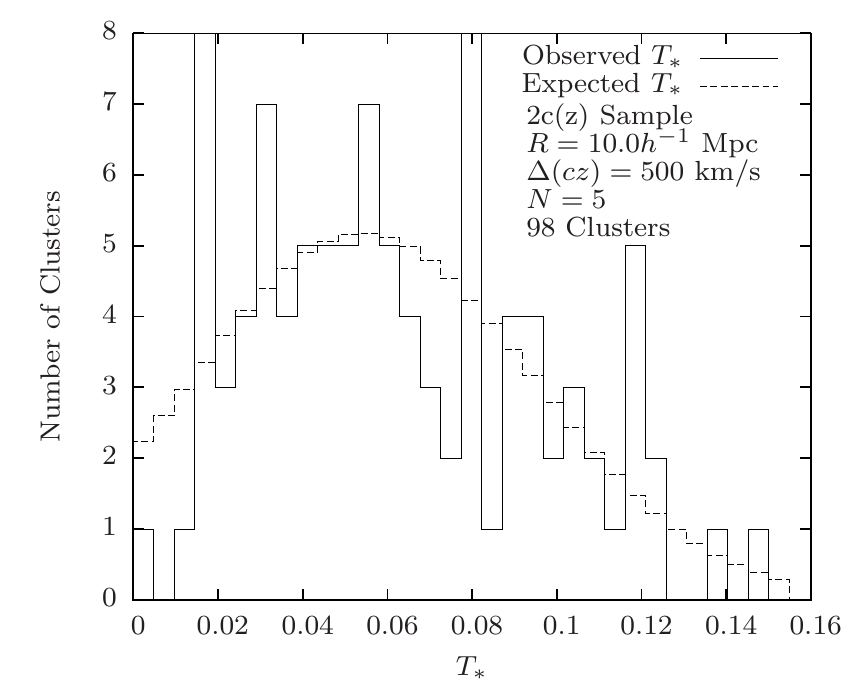}
\caption{
Observed and predicted histograms for $T_*$ for $10.0 h^{-1}$ Mpc cells with 500 km/s velocity dispersion illustrating the agreement with theory across different sample cuts. The solid line is the observed histogram and the dashed line is the expected histogram. Top left: 1a(z) sample; Top right: 1b(z) sample; Bottom left: 2b(z) sample; Bottom right: 2c(z) sample.
}
\label{fig-comp_Ts}
\end{center}
\end{figure*}

Comparing the histograms among the $a$, $b$ and $c$ samples, we see that the theory agrees with observations for different selection cuts and shows that it is a good description of clustering on a wide variety of scales. These may be as small as a $2.0 h^{-1}$ Mpc group or as large as a $20.0 h^{-1}$ Mpc supercluster. To illustrate this, we plot the histograms for different selection cuts at the cell size of $R=10.0 h^{-1}$ Mpc and $\Delta(cz)=500$ km/s with 5 galaxies in figure \ref{fig-comp_Ts} to demonstrate the agreement between theory and observations at these different scales.

\subsection{The Relationship Between $\eta$ and $\nu k_\tau$}
\begin{figure*}[tbp]
\begin{center}
\includegraphics[width=\floatwidth]{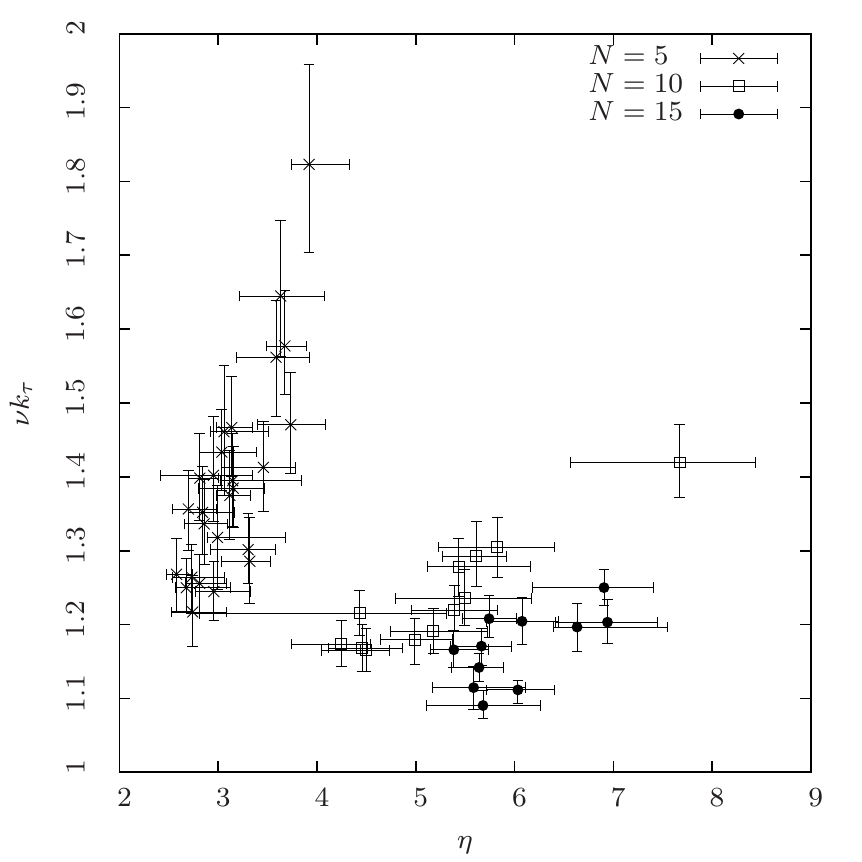}
\includegraphics[width=\floatwidth]{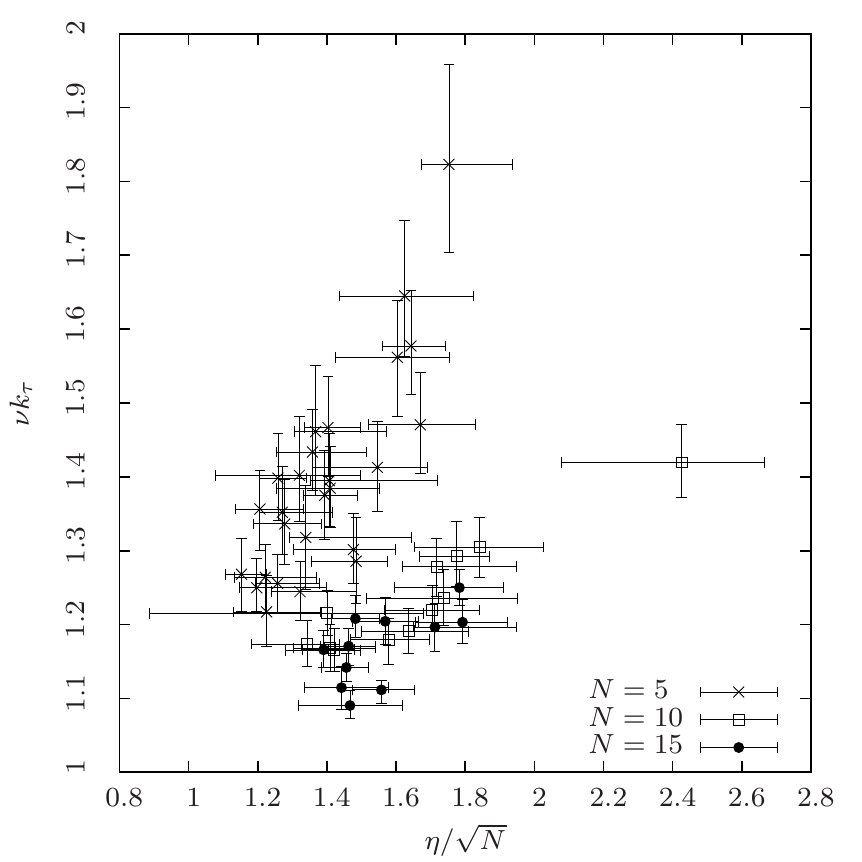}
\caption{
Left plot: Relation between $\nu k_\tau$ and $\eta$; Right plot: Relation between $\nu k_\tau$ and $\eta/\sqrt{N}$.
}
\label{fig-ktn}
\end{center}
\end{figure*}

From tables \ref{tab-summary-N5}, \ref{tab-summary-N10} and \ref{tab-summary-N15}, we see that the value of $\eta$ increases with $\nu k_\tau$ and suggests that, as expected, anisotropy in the shape of a cluster is correlated with anisotropy in the velocity distribution of a cluster. To illustrate this increase, we plot $\nu k_\tau$ against $\eta$ in figure \ref{fig-ktn} and show that this relation also depends on the number of galaxies in a cell.

When we plot $\nu k_\tau$ against $\eta/\sqrt{N}$, the points from instances with different $N$ line up with a weak correlation between $\nu k_\tau$ and $\eta/\sqrt{N}$. These data points indicate that $\nu k_\tau$ and $\eta/\sqrt{N}$ are close to their isotropic and uniform values. For an isotropic velocity distribution $\nu = \sqrt{3} \approx 1.73$ which is somewhat higher than the observed values of $\nu k_\tau$. This suggests that $k_\tau \lesssim 1$ if the clusters have an isotropic velocity distribution. This is generally reasonable since we are averaging over a sample of clusters that may have any orientation, and hence would on average be isotropic.

For a uniform spherical cell, $\eta/\sqrt{N} \approx 1.69$ which is also close but slightly less than the observed values, indicating that the identified clusters are close to, but not quite uniform. This suggests that the pairwise radial separation is larger than what is expected from a uniform cell, which may indicate the presence of internal structure, or a non-spherical shape for the cluster. A physical explanation for this difference is likely to be a combination of these factors and will depend on detailed models of galaxy clusters.

Comparing the probabilities for the $W_*$ histograms for different $N$, we see that the instances with denser cells tend to have higher probabilities. This is because the number of pairs in a cell scales as $N^2$, so that the larger number of pairs in a denser cell will lessen the relative contribution to $W_*$ from a close visual pair even though there are more of them. Less dense cells have significantly fewer pairs, and thus a very close pair will very easily dominate $W_*$ and result in a more pronounced tail.

\section{Conclusions}
\label{sec-conc}

This paper develops a method to find clusters of galaxies in a sky survey and estimates their scaled kinetic and correlation potential energies. Using the low redshift galaxies from DR7 of the SDSS, we identify a population of galaxy clusters and estimate the scaled energies of the cells that host them.

The distribution of scaled kinetic energy $T_*$, scaled correlation potential energy $W_*$ and correlation virial ratio $\psi$ generally agree with the theoretical predictions derived by \citet{2012ApJ...745...87Y} with free parameters that describe the 6-dimensional phase-space structure of a cluster. These parameters, $\eta$ which describes the shape of a cluster, $\nu$ which describes the velocity anisotropy and $k_\tau$ which describes the relation between the dynamical timescale and the crossing time, provide a new statistical method for estimating the radial distances, transverse velocities and the average mass of a galaxy.

In addition to these structural parameters, we have also introduced the parameters $\sigma_{W*}$, $\sigma_{T*}$ and $\sigma_\psi$ to model the statistical uncertainties in the estimated cluster energies. These parameters are the standard deviations of normal distributions that we convolve with the theoretical distribution of cluster energies and represent a combination of observational uncertainties and dynamical fluctuations. These may include uncertainties related to the detailed internal structure of a cluster that we cannot observe, and intrinsic fluctuations in the phase space configuration of clusters. The physical values of these parameters can eventually be determined from detailed models of clusters of galaxies, or suitably designed $N$-body simulations.

However, while the anisotropy parameters $\nu k_\tau$ and $\eta$ may be measured from a snapshot of the 6-dimensional phase space information of a sample of clusters from a simulation, the intrinsic fluctuations of a cluster about quasi-equilibrium require more information. A detailed analysis of such fluctuations is likely to require multiple snapshots of multiple clusters, at intervals considerably shorter than a crossing time. These requirements generally preclude the use of archived simulations since they are not archived with a sufficiently high time resolution. This is in order to track both the quasi-equilibrium average energies of a cluster and its fluctuations, and obtain the distribution of energies about quasi-equilibrium and its contributions to $\sigma_{T*}$, $\sigma_{W*}$ and $\sigma_\psi$.

Furthermore, the theory discussed in this paper and \citetalias{2012ApJ...745...87Y} suggests a strong connection between the structure of a cluster and the environment that it exists in. This suggests that the merger history and dynamics of a cluster is an important factor that determines its internal structure. Thus relating a high-resolution simulation to observations is considerably more complicated than a simple semi-analytic model, and in our case, even more so because we are interested in the detailed substructure of a cluster of galaxies.

Quantitatively, the observed distribution of $T_*$ agrees with the theoretical distribution convolved with a normal distribution, and is statistically significant at the 95\% level for most of the instances we have examined. However, the observed distributions for $W_*$ and $\psi$ have a long tail that does not agree with theory. This tail is likely to be caused by the presence of a population of merging galaxies. These merging galaxies are very close to each other, and contribute to a very negative $W_*$. These pairs are likely to eventually become a single galaxy, and should be modeled as a single extended galaxy~\citep{2010arXiv1011.0176Y}. 

In order to account for this long tail, we need to identify the merging pairs and consider them as a single extended galaxy. We do not do so in this paper because such an analysis warrants a much more detailed treatment to deal adequately with the merger classification methods and would be better addressed in a separate paper.

We also find that the quasi-equilibrium theory of galaxy clusters holds for a large range of scales. These range from small groups in cells of $2.0 h^{-1}$ Mpc radius, to large supercluster scale structures in cells of $20.0 h^{-1}$ Mpc radius. This agrees with the result that the GQED agrees very well with the counts-in-cells distribution at a wide variety of scales~\citep{2011ApJ...729..123Y,2005ApJ...626..795S}.

By analyzing different samples of galaxies and clusters, we have also found that $\nu k_\tau$ is weakly correlated with $\eta/\sqrt{N}$ and indicates that the velocity anisotropy and position anisotropy of a cluster are weakly correlated. However a more general result is that $\nu k_\tau$ and $\eta/\sqrt{N}$ are close to the isotropic and uniform values which show that, on average, clusters are approximately isotropic, and are close to, but not quite uniform collections of galaxies.

We conclude that the analysis here suggests that the quasi-equilibrium theory in \citetalias{2012ApJ...745...87Y} is a good description of galaxy clustering when the uncertainties and fluctuations in the cluster kinetic and correlation potential energies are incorporated. While some of these uncertainties serve to broaden the distribution, the spatial and velocity anisotropy parameters $\eta$ and $\nu$ may provide further insights to the internal structure of galaxy clusters on a statistical basis. These parameters, and the intrinsic fluctuations around quasi-equilibrium may be measured in suitably designed $N$-body simulations. Using the conclusions in this paper, we are currently working on a subsequent paper that will discuss simulations.

\acknowledgements

We thank the referee for very helpful comments that has helped to make this paper more self-contained, especially in the discussion of quasi-equilibrium.

Funding for the creation and distribution of the SDSS Archive is provided by the Alfred P. Sloan Foundation, the Participating Institutions, the National Aeronautics and Space Administration, the National Science Foundation, the US Department of Energy, the Japanese Monbukagakusho, and the Max Planck Society. The SDSS Web site is at \url{http://www.sdss.org}.

The SDSS is managed by the Astrophysical Research Consortium (ARC) for the Participating Institutions. The Participating Institutions are the University of Chicago, Fermilab, the Institute for Advanced Study, the Japan Participation Group, Johns Hopkins University, the Korean Scientist Group, Los Alamos National Laboratory, the Max-Planck-Institute for Astronomy (MPIA), the Max-Planck-Institute for Astrophysics (MPA), New Mexico State University, University of Pittsburgh, Princeton University, the US Naval Observatory, and the University of Washington.

\bibliography{paper}
\end{document}